# Are female scientists less inclined to publish alone? The gender solo research gap

**Marek Kwiek**[1] · **Wojciech Roszka**[2]

**Abstract**
In solo research, scientists compete individually for prestige, sending clear signals about their research ability, avoiding problems in credit allocation, and reducing conflicts about authorship. We examine to what extent male and female scientists differ in their use of solo publishing across various dimensions. This research is the first to comprehensively study the "gender solo research gap" among all internationally visible scientists within a whole national higher education system. We examine the gap through mean "individual solo publishing rates" found in "individual publication portfolios" constructed for each Polish university professor. We use the practical significance/statistical significance difference (based on the effect-size *r* coefficient) and our analyses indicate that while some gender differences are statistically significant, they have no practical significance. Using a partial effects of fractional logistic regression approach, we estimate the probability of conducting solo research. In none of the models does gender explain the variability of the individual solo publishing rate. The strongest predictor of individual solo publishing rate is the average team size, publishing in STEM fields negatively affects the rate, publishing in male-dominated disciplines positively affects it, and the influence of international collaboration is negative. The gender solo research gap in Poland is much weaker than expected: within a more general trend toward team research and international research, gender differences in solo research are much weaker and less relevant than initially assumed. We use our unique biographical, administrative, publication, and citation database ("Polish Science Observatory") with metadata on all Polish scientists present in Scopus ($N=25{,}463$) and their 158,743 Scopus-indexed articles published in 2009–2018, including 18,900 solo articles.

✉ Marek Kwiek
kwiekm@amu.edu.pl

Wojciech Roszka
wojciech.roszka@ue.poznan.pl

[1] Institute for Advanced Studies in Social Sciences and Humanities, UNESCO Chair in Institutional Research and Higher Education Policy, Adam Mickiewicz University of Poznan, Poznan, Poland

[2] Poznan University of Economics and Business, Poznan, Poland

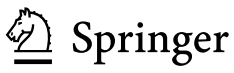

## Introduction

In the highly competitive global science, publications are a major determinant of successful academic careers (Stephan, 2012). However, publications represent various types of authorship, with the major distinction between solo and team research. Single-authored publications, characterized as being doomed to extinction for the past three decades in research literature (spanning scientometrics, sociology of science, economics of science, and higher education research) but still continuing to exist, ask for analytical attention as a special mode of academic knowledge production (Kuld & O'Hagan, 2018). Our interest here is in what we term "the gender solo research gap," or differences in using solo publications by male and female scientists (best visible through mean individual publishing solo rate: fractions of single-authored publications among all publications in their ten-year publication portfolios), and its impact on academic careers. Solo publications reflect the traditional vision of knowledge in which individual scientists, rather than their teams, contribute to scientific discoveries. Although this perspective has been changing, with ever greater emphasis on team research, solo research continues to exist, albeit with different roles in different disciplines (West et al., 2013).

In an academic world in which gender gaps in research funding (Cruz-Castro & Sanz-Menéndez, 2019; Van den Besselaar & Sandström, 2016), promotion and tenure (Rivera, 2017; Weisshaar, 2017) and salary (Ceci et al., 2014; Diezmann & Grieshaber, 2019) prevail, solo research is the only publishing mode in which there seems to be no ambiguity in credit allocation for research achievements, no errors in signals about scientists' research abilities, and no "biased credit attribution" (Sarsons et al., 2020: 31). In solo research, signals are not "noisy": men and women are treated similarly as sole authors and "receive the same amount of credit" (Sarsons et al., 2020: 32). Thus, solo research is a special case of credit allocation in science and it may play a special role in tackling gender discrimination, especially in terms of tenure. As recently shown in the case of economics, women face an enormous penalty for collaboration: while men get the same amount of credit for collaborative and solo research, women get essentially zero credit in tenure decisions if they collaborate with men (Wang & Barabási, 2021: 144–146). In most mixed-sex authorship configurations, female scientists are gaining less credit than deserved and exclusively same-sex publications tend to keep female scientists in female ghettos, limiting their academic and professional networks and hence their academic impact.

Solo research is thus a special case of authorship and publishing strategy which deserve to be studied in more depth because of its unbiased signaling of scientists' ability, credibility and independence. The role of solo research in academic careers seems potentially very important. However, solo research has not been examined from the perspective of sex differences, as opposed to numerous studies of collaborative research: are there noticeable and statistically significant differences between men and women in their use of solo research as part of their publishing strategies? Do the differences in publishing and collaboration patterns hold across major dimensions of academic careers (such as age, institutional type, academic seniority, and academic disciplines)?

We first briefly summarize gender gaps in science, adding the gender solo research gap to a long list of existing gaps; then we summarize the existing research on solo research in various contexts; and finally we move on to examine the relationship between publishing solo, gender and selected individual and institutional variables relevant for academic careers.

In the early twentieth century, the publication author was simply the single author. However, individual science gradually changed into team science over the century (Larivière et al., 2015; Wuchty et al., 2007), with exponential growth of co-authored publications accompanied by expectations that solo research would disappear (Price, 1963). The most characteristic tendency in publishing in the twenty-first century is the "intensifying scientific collaboration" (Gläznel, 2002: 461). Currently, assigning credit to, and receiving recognition from, collaborative research is still an unresolved problem; this has been highlighted extensively in the past three decades (Allen et al., 2014; Bridgstock, 1991; Endersby, 1996; Sarsons, 2017). Current thinking about science and its progress deeply rooted as it is in the history of science, with the sole author on the science pedestal for centuries (Shapin, 1991) has not caught up with daily practices in science in which team publishing predominates; in such daily practice, there is an increasing share of collaborative research in global science, and the average team size is increasing both in natural (Huang, 2015) and social sciences (Henriksen, 2016). Consequently, with the increasing division of labor, specialization, and hierarchy in larger teams, it is difficult to clearly identify who should be given credit as the main "authors" of the paper (Jabbehdari & Walsh, 2017: 2).

Our focus is on gender differences in solo research from the macro-level perspective of a single national system. Our dataset comprises all scientists with doctoral degrees employed full time in the higher education sector and all their Scopus-indexed publications, including all solo publications, in all academic disciplines. Our focus is on how male and female scientists of various biological ages, academic positions, and institutional types make use of, and benefit from, solo publishing.

Solo publishing, including gender differences in solo publishing, has not been studied comprehensively in terms of whole national systems, all age groups, academic positions, and disciplines. Kuld and O'Hagan (2018) examined 175,000 articles in 255 top journals in economics; Vafeas (2010) studied 25 accounting and finance journals over a five-year period; Nabout et al. (2015) studied four sub-areas of biology; and Ghiasi et al. (2019) studied 1.18 million solo articles published in 2008–2017 and indexed in the Web of Science. Generally, solo research has appeared in the margins of the studies of multi-authored papers. The sub-issues related to the gender solo research gap can be characterized as follows: (1) solo research and career stages (e.g., early career, mid-career, and established scientists); (2) solo research and disciplines (or cross-disciplinary differences); (3) solo research and institutions/institutional types (or cross-institutional differences); (4) solo research and academic rewards (e.g., tenure, research grants, and academic recognition); and finally, (5) solo research and disciplinary, institutional, and national academic cultures (and their changes over time). However, our empirically-driven focus is limited by our data because we do not have the reliable data on research grants, new hires, and changes in academic cultures over time.

In this study, gender has been unambiguously defined for all scientists, and all solo articles produced in a national system in 2009–2018 and indexed in Scopus have been gender classified. Using our newly constructed "Polish Science Observatory" database (see Kwiek & Roszka, 2021a, 2021b), we examined all male and female scientists with their biographical and administrative histories, including their biological age and clearly defined academic positions, as well as locations of their institutions and their dominant disciplines; we then investigated all their research articles, whether published solo or in teams. Based on previous global research literature, we assumed gender differences in solo research to be significant, and we examined them through "individual publication portfolios" specifically constructed for each scientist ($N=25{,}463$ scientists, all with doctorates, with 158,743 articles published in 2009–2018, including 18,900 solo articles; in our sample, 2887 female scientists authored 6119 solo articles, and 4871 male scientists authored 12,781 solo articles). For contextual purposes, we also examined our parallel "OECD Science Observatory" database of all (gender-defined) scientists and all (gender-classified) articles indexed in Scopus from 1674 research-involved institutions from 38 OECD economies in the same period (2009–2018), encompassing 11,087,392 scientists and 27.4 million publications.

## Key literature

### The context: the gender gaps in science

The gender solo research gap accompanies other gender gaps in science. Systematically reviewing research literature on male–female differences in science, we have identified specific gender gaps in 15 areas; to which we add a new one: gender solo research gap. These gender gaps refer to productivity, self-citations, international collaboration, mobility, professional networks, research funding, academic time distribution, academic role orientation, disciplines, methods, and research agendas (defined here as "input-related gender gaps"), as well as citations, group work recognition, tenure, and salaries (defined here as "output-related gender gaps"; see Fig. 1).

Input-related gaps accompany the process of research production; output-related gaps, in contrast, accompany the processes of assessment and reward in science (and assessment and rewarding of scientists, its producers). Among the rewards in science, we identify differences in how male and female scientists are cited, how their role in collaboration is assessed, how they obtain their tenure, and what salaries they receive. Good examples of input versus output-related gender gaps are self-citations for the former gap class and citations for the latter. Female scientists tend to self-cite less (King et al., 2017; Maliniak et al., 2013) on the input side, but they also tend to be cited less (Ghiasi et al., 2018; Potthof and Zimmermann, 2017) on the output side. Instead of providing a wide panorama of gender gaps in science and how they operate, a list of the 16 gender gaps with a few representative studies is presented in Table 1. However, on top of that, gender gaps in science function within much larger gender socioeconomic gaps, with the former being clearly linked to the latter; for instance, gender relations in universities cannot be easily disassociated from gender relations in societies at large (including balancing work and family roles, the role of religion and patriarchy in societies, etc.; see a comprehensive account in Lindsay, 2011).

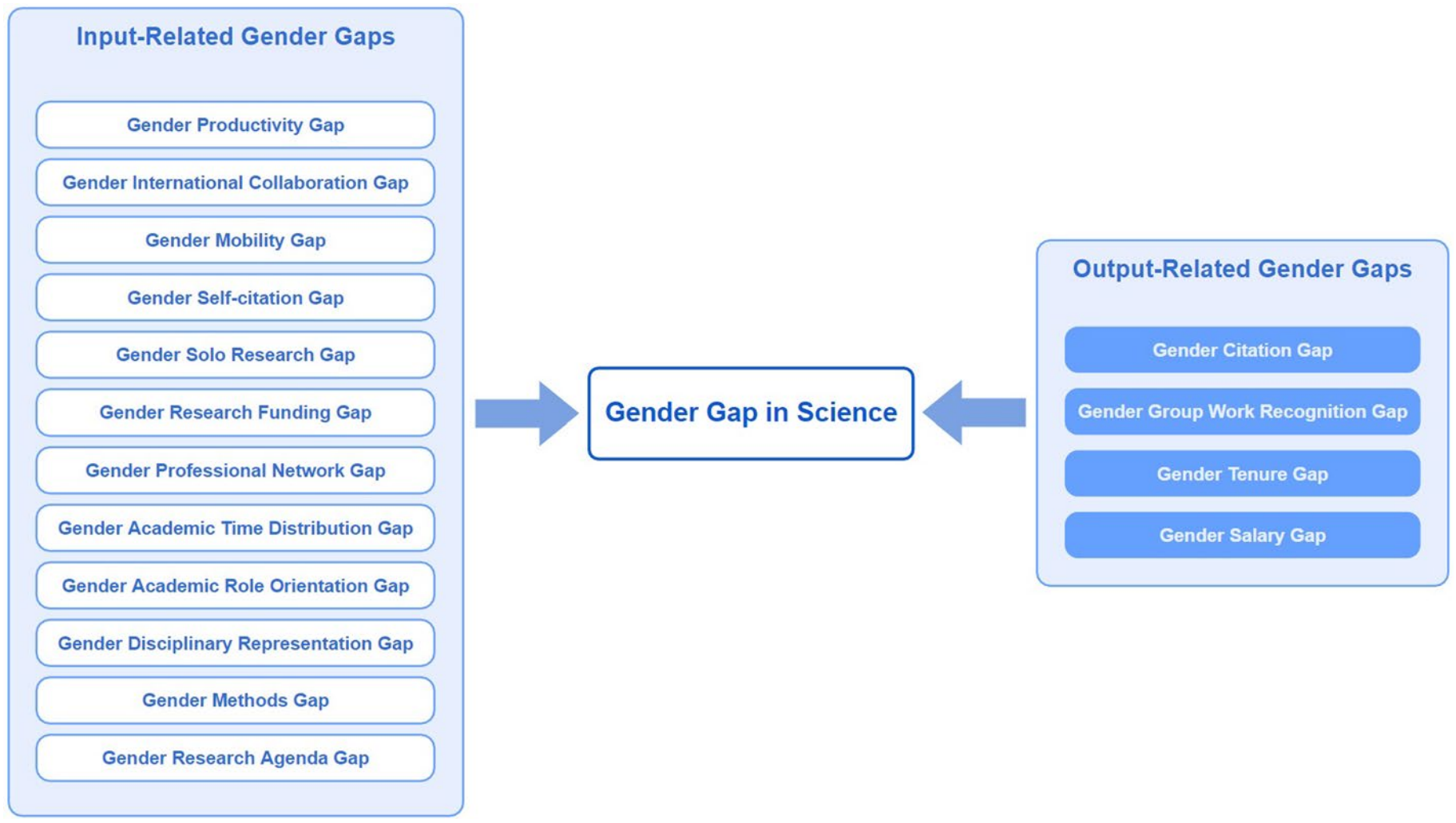

**Fig. 1** Gender gaps in science: a classification

**Table 1** Gender gaps in science: brief literature review

| Type | Simple explanation | Selected literature gaps examined |
|---|---|---|
| Gender *productivity* gap | Female scientists are less productive than male scientists are | Larivière et al. (2011, 2013), Nielsen (2016), van den Besselaar and Sandström (2016), Mihaljević-Brandt et al. (2016), van den Besselaar and Sandström (2017), Abramo et al. (2019), Huang et al. (2020) |
| Gender *collaboration* gap | Female scientists are less involved in (especially international) research collaboration than male scientists are | Bozeman et al. (2012), Larivière et al. (2013), Abramo et al. (2013), Vabø et al. (2014), Fell and König (2016), Nielsen (2016), Fox et al. (2017), Aksnes et al. (2019), Maddi et al. (2019), Kwiek (2020), Kwiek and Roszka (2021a), Fox (2020) |
| Gender *mobility* gap | Female scientists are less involved in (especially international) physical mobility than male scientists are | Ackers (2008), Frehill and Zippel (2006), Jöns (2011), Zippel (2017), Uhly et al. (2017) |
| Gender *self-citation* gap | Female scientists cite themselves less often than male scientists do | Hutson (2006), Maliniak et al. (2013), King et al. (2017), Mishra et al. (2018) |
| Gender *solo research* gap | Female scientists are less involved in publishing alone than male scientists are | West et al. (2013), Walker (2019), Sarsons et al. (2020) |
| Gender *research funding* gap | Female scientists are awarded smaller research grants or receive them less often compared with male scientists | Larivière et al. (2011), van den Besselaar and Sandström (2015), van den Besselaar and Sandström (2017), Cruz-Castro and Sanz-Menéndez (2019). Opposing evidence: Marsh et al. (2009) |
| Gender *professional network* gap | Female scientists have narrower (and less international) formal and informal networks of collaborators than male scientists do | Feeney and Bernal (2010), Van den Brink and Benschop (2013), Kegen (2013), Clauset et al. (2015), Greguletz (2018), Halevi (2019), Heffernan (2020) |
| Gender *academic time distribution* gap | Female scientists spend more time on teaching and male scientists more time on research | Toutkoushian and Bellas (1999), Cummings and Finkelstein (2012), Leišytė and Hosch-Dayican (2017), Goastellec and Vaira (2017), Kwiek (2019) |
| Gender *academic role orientation* gap | Female scientists are less research-oriented and more teaching-oriented than male scientists are | Miller and Chamberlin (2000), Cummings and Finkelstein (2012), Leišytė and Hosch-Dayican (2017), Goastellec and Vaira (2017), Kwiek (2019) |
| Gender *disciplinary representation* gap | Female scientists are underrepresented in large parts of STEM fields compared with male scientists | Ceci and Williams (2011), Shapiro and Williams (2011), Ceci et al. (2014), Avolio et al. (2020) |
| Gender *methods* gap | Female scientists use quantitative methods less often and qualitative methods more often than male scientists do | Thelwall et al. (2019), Key and Sumner (2019) |

Women are massively involved in research, but gender gaps continue to exist, possibly widening in some areas. Polish women constituted 41.50% of university professors of all ranks, with at least a doctoral degree in our sample and 47.56% of the entire full-time academic workforce in 2020 (Statistics Poland, 2021: 126). A mechanism that may contribute to widening rather than closing gender gaps in science is "cumulative disadvantage", or the "accumulation of failures" (Cole, 1979: 78), representing the reverse of Merton's (1968) "cumulative advantage." Processes of accumulative advantage for male scientists may be accompanied by processes of accumulative disadvantage for female scientists in which the negative impact of some or all gender gaps combined build up over time (with the "self-reinforcing dynamic" ever stronger; van den Besselaar & Sandström, 2017: 14). As the rich (in citations, publications, international collaboration, mobility, funding, professional networks, research time, tenure, recognition, etc.) become richer, the poor here, female scientists embedded in gender gap–ridden academic environment become relatively poorer.

## Solo research and the individual authorship decision

Research literature tends to show the future of solo research in dramatic terms; while the "decline" of solo publications has been discussed for several decades, more recently, the "extinction of the single-authored paper" has appeared "imminent" (in ecological research; Barlow et al., 2017). Moreover, "the demise of the 'lone star'" as the author of solo research is discussed, even though the results from research indicate "relative decline" (in economics; Kuld and O'Hagan, 2018). Solo research is conceptualized as a "vanishing breed," particularly in life sciences (Allen et al., 2014), and the "demise of single-authored publications" is reported in computer science (Ryu, 2020). Apart from a long list of factors explaining why solo research is disappearing, two technical factors are important to highlight as follows: the tendency of supervisors to co-author with their students and doctoral students and a shift from informal to formal collaboration in which scientists are making sure that their contributions are visible (Henriksen, 2016). As a paper on Austrian life science postdocs summarizes its qualitative findings, "nearly every act of technical or epistemic support constitutes an implicit exchange relationship; publication credits are received for the time and knowledge invested" (Fochler et al., 2016: 193).

In practical terms, solo research stems from voluntarily making individual authorship decisions. Individual scientists make consequential authorship decisions about how to follow in their research, and choosing solo publications is one of the options (as is choosing same-sex or mixed-sex collaborators, McDowell et al., 2006; on the gender-based homophily in research, or men collaborating with men and women collaborating with women, see Kwiek & Roszka, 2021b). The authorship decision is important because "it is likely to affect the project's quality, efficiency of execution, and exposure, as well as the amount of credit an author receives following its eventual publication" (Vafeas, 2010: 332). The results of many individual authorship decisions accumulate over time, accompanying lifetime academic careers.

**Table 1** (continued)

| Type | Simple explanation | Selected literature gaps examined |
|---|---|---|
| Gender *research agenda* gap | Female scientists study different research topics than male scientists do | Key and Sumner (2019), Thelwall et al. (2019), Santos et al. (2020) |
| Gender *citation* gap | Female scientists are less cited than male scientists are | Aksnes et al. (2011), Maliniak et al. (2013), Ghiasi et al. (2015), Abramo et al. (2015), Potthof and Zimmermann (2017), van den Besselaar and Sandström (2017), Ghiasi et al. (2018), Lerchenmueller et al. (2019), Maddi et al. (2019), Huang et al. (2020), Madison and Fahlman (2020), Thelwall (2020) |
| Gender *group work recognition* gap | Female scientists receive less deserved recognition (or less deserved credit) for their collaborative publications than male scientists do | Heffner (1979), Sarsons (2017), Sarsons et al. (2020) |
| Gender *tenure* gap | Female scientists are less often promoted to tenure than male scientists are despite equal achievements | McDowell and Smith (1992), Abramo et al. (2015), Fell and König (2016), Weisshaar (2017), Rivera (2017), Diezmann and Grieshaber (2019), Sarsons et al. (2020). Opposing evidence: Madison and Fahlman (2020) |
| Gender *salary* gap | Female scientists have lower salaries in the same positions in academia than male scientists do | Fox 1985), Barbezat and Hughes (2005), Ward and Sloane (2000), Ceci et al. (2014), Kwiek (2018a) |

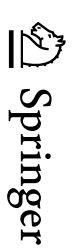

Authorship decisions may bear on tenure decisions and the availability of external research grants from national research councils. Major research funding agencies may favor not only publications in top international journals but also publications written in international collaboration (Kwiek, 2015), following the global and European "internationalization imperative" (Ackers, 2008) in research policies and a generally assumed link between research internationalization and productivity (Kwiek, 2016; a global exception to a positive role of research internationalization in promotion, tenure, salaries, and research grants being the United States, see Cummings & Finkelstein, 2012). Therefore, authorship decisions need to be "intelligent" (Vafeas, 2010: 333) and "strategic" (Jeong et al., 2011: 968). The major discrete choice among the collaboration modes is between solo research and team research and then between the various types of team research. Women are reported to be significantly underrepresented not only as first and last authors of publications (Walker, 2019) but also as authors of solo research (Sarsons et al., 2020; Walker, 2019; West et al., 2013).

Individual authorship decisions may be critical for individual academic careers; however, the implications go far beyond individual scientists and reach the aggregated levels of institutions, disciplines, and national systems. For instance, Poland has the lowest level of international collaboration in research among all 27 European Union countries and the second highest levels in solo publishing at 36.0% and 12.1%, respectively (Scopus, 2021; see Kwiek, 2021b). The publishing patterns at aggregated levels, such as those for an institution or a country, depend entirely on individual decisions of thousands of scientists who are willing to publish solo or in institutional, national, or international collaboration.

Collaboration, as opposed to solo publishing, involves compromise and tends to reduce risk taking (Hudson, 1996: 157; Kuld and O'Hagan, 2018: 1221). However, collaboration can result in information overload, unclear responsibilities, and communication issues collectively known as "coordination costs" (Olechnicka et al., 2019: 111). Scientists make consequential decisions not only about where to publish (within a steep hierarchical order of academic journals; see the "prestige-maximization" role of top journals in Kwiek, 2021a) but also about whether to publish solo or to collaborate based on the available resources, research environment, and trade-offs among alternative modes of collaboration (Jeong et al., 2014: 521).

The gender solo research gap can be examined along two lines: sex differences in selecting solo publishing, and sex differences in the intensity of publishing solo (only for scientists involved in this publishing type).

### Solo research and academic reputation

Academic reputation comes almost exclusively from publications (Stephan, 2012), just as social stratification in science is largely publication based (Kwiek, 2019). It seems to be closely linked not only to team publications but also to solo publications. The disciplines studied in the literature in the context of solo research include accounting (Rutledge & Karim, 2009), mathematics (Mihaljević-Brandt et al., 2016), social sciences and humanities (Larivière et al., 2006), political sciences (Fisher et al., 1998), and life sciences (Fochler et al., 2016; Müller, 2012; Müller & Kenney, 2014). Specifically, the link between academic reputation and solo research pertains to highly prolific and highly cited authors (Vafeas, 2010): a certain minimum amount of solo publications may be needed to belong to the global research elite (Kwiek, 2016), and solo publications for this specific layer of top scientists may be strategically located in highly prestigious journals.

As a study of accounting shows, prolific authors in accounting literature become more productive and produce longer articles using extensive collaboration. However, such prolific authors "appear to decrease the number of co-authors on their higher quality publications, possibly to increase the quality of their reputation" (Rutledge & Karim, 2009: 130). Interestingly for our research, regression results suggest that productive authors' publications that use fewer co-authors are more likely to appear in journals with a greater impact on the literature (Rutledge & Karim, 2009: 133). Scientists are reported to be more likely to publish solo research if they are affiliated with universities located higher in rankings, when the expected amount of work (proxied by the article's length) is small, and if the article is conceptual rather than empirical (Vafeas, 2010: 340–341). The university rank is significantly related to the likelihood of single authorship, with authors from highly ranked institutions "having the training and resources to be more self-sufficient in conducting their research" (Vafeas, 2010: 341). There may be a tendency of highly cited scientists to publish their solo research, rare as it is, in top journals in their disciplines. Although we will not compare highly prolific male and female scientists in this research, we will analyze male and female scientists located in two institutional types ("Solo research, academic discipline and institutional research intensity by gender ", "A modeling approach: a fractional logit regression model" Sections): ten research-intensive universities and the remaining institutions in the system.

## Solo research, credit allocation, and authorship claims

Solo research avoids problems in credit allocation for publication (Sarsons, 2017; Sarsons et al., 2020), and it may reduce possible conflicts about authorship (Barlow et al., 2017). Academic publications are key to the individual futures of young scientists, men and women alike, especially when large cohorts of postdoctoral researchers seek permanent jobs (confirming the role of "cohort effects"; Stephan, 2012: 174–176). Young scientists fight for academic survival in a rapidly changing academic world in which doctoral students are already expected to publish, and postdocs are expected to publish extensively; such expectations were lower in the late twentieth century. High-quality research performance matters because, as Stephan (2012: 149) comments, "no output, no funding". However, cohort also matters: in this case, what is important is the current global abundance of postdoctoral researchers and the scarcity of academic employment opportunities for them. The supply of highly able doctorates exceeds the demand for postdoctoral opportunities, not to mention permanent jobs. A series of studies based on in-depth analyses of interviews with postdocs in life science about their academic career rationales (Fochler et al., 2016; Müller, 2012; Müller & Kenney, 2014) highlight growing tensions related to the choice of preferred working style and publishing pattern in their day-to-day practices. In a hypercompetitive academic environment in which the supply of postdocs in life sciences (as in other disciplines) is much higher than the demand for candidates for full-time academic jobs, young scientists with doctorates have to ensure first authorship (or solo authorship) for their publications if they want to send clear signals to the academic labor market about their outstanding research abilities.

Publication, and hence, the question of authorship, is pivotal in negotiations about collaboration in ongoing research. The postdoc's choice is often to work individually to avoid possible authorship conflicts; postdocs are reported to use those collaborative opportunities that "do not pose a threat to individual authorship claims" (Müller, 2012: 291). In fast-growing, highly internationalized, and highly competitive research fields in which science is expected to be highly collaborative young scientists, paradoxically, may choose individualized modes of working and publishing. The reason is simple: in solo (or, to some extent, first-author papers), it is clear where credits for publication go. Strategic thinking may involve considering solo research more strongly in one's 30s than in one's 40s although certainly not in all disciplines. In Europe, with its highly prestigious, multibillion-euro European Research Council financing thousands of scientists, publications co-authored with dissertation supervisors tend not to count in competitions for early career researchers.

While publishing in co-authorships is safer (the risk of openly hostile criticism is reduced, and the responsibility for errors is divided between all co-authors), it may not suffice to obtain a permanent job, or in some systems, to keep it. In most disciplines, first-author publications are as powerful signals of individual research ability as solo publications are. Although Price (1963) expected that "by 1980 the single-author paper will be extinct," Abt (2007: 358) was right when he claimed that single-authored papers would not disappear soon because "there are some projects that do not require teams and some authors who prefer to work individually."

On top of that, benefits from team research always needs to be juxtaposed with the costs and risks of it which may differ by gender. Management costs for team research, with more people, institutions, and countries involved, tend to be higher. Specifically, transaction costs (Georghiou, 1998) and coordination costs (Cummings & Kiesler, 2007) are higher for international research collaboration and they may be higher for female than for male scientists. Women may be more negatively affected by physical mobility requirements in international collaboration (Ackers, 2008; Zippel, 2017). In team research, there is a trade-off between increased publication numbers and access to research funds and the need to minimize transaction costs (Landry & Amara, 1998).

## Solo research and academic discipline, age, and seniority

Solo research is differently distributed across disciplines, which exhibit distinct dominant collaborative practices; consequently, one might not expect a direct comparison between the team size of papers in mathematics versus physics and astronomy (Huang, 2015), the former being a low-author field and the latter a high-author field on average. Average team size is highly differentiated across disciplines (Larivière et al., 2015), and in disciplines that are heavily solo research dominated, such as some in the arts and humanities and in social sciences, authorship credit is usually attributed to a single author (Endersby, 1996: 381). As reported for Canada, "in the humanities and literature, formal collaboration based on co-authorship is a marginal phenomenon" (Larivière et al., 2006: 531). As demonstrated for seven major academic institutions in Israel, the more theoretical the research, the higher the probability of the paper being single authored (Farber, 2005: 65), as confirmed also for political science journals (Fisher et al., 1998: 855). There are also significant cross-institutional differences in the numbers and shares of solo publications, with mathematics identified as a discipline with a remarkably higher number of single-authored papers across all seven Israeli institutions (Farber, 2005: 64); and cross-journal differences within disciplines, as in political science (Fisher et al., 1998). We examine gender differences by discipline in "Solo research, academic discipline and institutional research intensity by gender " Section.

While the general opposition between solo and team research is analytically useful, it does not allow telling the whole story, especially the story of ongoing evolution in dominant authorship types by discipline. In some disciplines, the historical change in the past quarter of a century is away from solo and toward team publications, while in others, it is away from two-author and toward three-author publications. The trends for two-author and three-author publications may not be the same, and the trends for two-author and 10-author publications, both being generally team publications, may differ substantially. “Small-group collaborations” may have different dynamics by discipline and over time than “large-group collaborations” do. Consequently, “small-group collaborations” in basic sciences as stochastic processes differ from “large-group collaborations” as staggered plateaus (Huang, 2015: 2141–2146). In different disciplines, there are different authorship types dominating at one time, and research collaboration in different disciplines may go through the same stages, but with a delay, at different times (Huang, 2015: 2146).

Solo research emerges from literature as strongly related to age and academic seniority which we study empirically in "Solo research, biological age, and gender" Section. Junior faculty are reported to show a higher propensity to publish solo research than senior faculty are, with two independent explanations for this phenomenon: first, junior faculty use single authorship as signals about their ability to perform independent research; and second, junior faculty are sole authors in papers coming out of their doctoral theses (Vafeas, 2010: 341), in the Polish case, when they are about 30–35 years old. As Kuld and O’Hagan (2018) have shown for journals in economics, younger scientists publish significantly more solo-authored papers than older scientists do. In economics, over 20% of all articles in top journals are solo authored, and in many cases, solo-authored articles have citation counts as high as or higher than citations for papers with multiple authors (Kuld & O’Hagan, 2018: 1223). Solo research may also suggest a higher degree of research independence and credibility, and it may be useful in the academic job marketplace at the postdoctoral level.

## Research questions and hypotheses

Our four research questions and hypotheses are shown in Table 2, together with support received in the Results section of the paper. They are strongly linked to findings from previous studies analyzed above in the context of solo research in general (authorship decisions; academic reputation; credit allocation) and solo research and gender.

We want to examine sex-related differences in solo publishing as a specific publishing strategy and assess the scope and possible impact of these differences within a national context. The research questions we follow are also closely related to the data at our disposal: in this empirically-driven paper, we are unable to examine the questions for which no data are available, such as, for instance, wider changes of disciplinary, institutional and national academic cultures toward gender equality ongoing in Polish universities; or the differentiated impact of a decade of higher education reforms on beliefs and attitudes regarding publishing and collaboration patterns of male and female scientists for which reliable data would come from national surveys. The hypotheses pertain to academic disciplines (H1), institutional research intensity (H2), and biological age (H3). Finally, we are also studying the relationship between the propensity to publish solo and gender (H4) using fractional logit regression models.

**Table 2** Research hypotheses and results (summary)

| Research Question | Hypothesis | Support |
|---|---|---|
| RQ1. What is the relationship between publishing solo, gender, and academic disciplines? | H1: Academic disciplines: We expect that female scientists will exhibit lower individual publishing solo rates than male scientists across all academic disciplines | Not supported: gender differences with no practical significance |
| RQ2. What is the relationship between publishing solo, gender, and institutional research intensity? | H2: Institutional research intensity: We expect that female scientists will exhibit higher individual publishing solo rates than male scientists in research-intensive institutions and lower individual publishing solo rates than male scientists in institutions less involved in research | Not supported: gender differences with no practical significance |
| RQ3. What is the relationship between publishing solo, gender, and biological age? | H3: Biological age: We expect that younger male and female scientists will exhibit higher individual publishing solo rates than older male and female scientists | Not supported: gender differences with no practical significance |
| RQ4. What is the relationship between the propensity to publish solo and gender? | H4: Solo publication propensity: We expect that being a female scientist decreases the propensity to conduct solo research or the individual publishing solo rate (in fractional logit regression models) | Mixed results: marginal influence of gender |

## Data and methods

### Dataset

We used "The Polish Science Observatory" database (see Kwiek & Roszka, 2021a: 4–6 for its description). Two large databases were merged. An official register of all Polish academic scientists was merged with the Scopus database of author's names/IDs and publications, with all metadata available. The merger of the biographical and administrative dataset (The Polish Science) with the publication and citation database (Scopus) used both probabilistic record linkage and deterministic record linkage. Database I comprised 99,535 scientists. The data used were demographic (gender and date of birth) and professional (the highest degree awarded; award date of PhD, habilitation, and full professorship; and primary institutional affiliation), with each scientist identified by a unique ID. Database II included 169,775 names from 85 institutions whose publications for the decade analyzed (2009–2018) were included in the Scopus database and 384,736 Scopus-indexed publications. Probabilistic methods of data integration were used (see Herzog et al., 2007; Enamorado et al., 2019). An integrated database used in this research finally included 32,937 unique authors of publications of various types, including 25,463 authors of journal articles. Our dataset had 7758 solo scientists (i.e., scientists with at least a single solo article, 4871 male and 2887 female scientists) and 19,252 solo articles. In the "Observatory" database, every Polish academic scientist with a doctoral degree is characterized by the dominant discipline (one of 27 ASJC general disciplines, ASJC is All Science Journal Classification in Scopus). Consequently, we had a clearly defined gender, biological age, and dominant discipline for every scientist, along with all their solo and team publications, as well as the distribution of female and male scientists in every discipline. The dominant disciplines, individual publication portfolios, gender composition of disciplines, and average publication prestige were constructed for the decade of 2009–2018.

### Methods

As in our previous research on gender disparities in international research collaboration (Kwiek & Roszka, 2021a) and on gender-based homophily in research or man-woman collaboration patterns (Kwiek & Roszka, 2021b), also here every Polish scientist represented in our integrated database was ascribed to one of 27 ASJC disciplines at the two-digit level (following Abramo, Aksnes, & D'Angelo, 2020). A paper can have one or multiple disciplinary classifications (see the ASJC discipline codes used, as described in Table 3). The dominant discipline for each scientist is the mode for each of them: the most frequently occurring value (when no single mode occurred, the dominant discipline was randomly selected). Consequently, all Polish scientists with Scopus-indexed articles were defined by their gender, discipline, as well as their publications included in their individual publication portfolios. Every ASJC discipline in Poland represents proportions of male and female scientists: they are either male-dominated or female-dominated disciplines. However, GEN, NEURO, and NURS disciplines did not meet an arbitrary minimum threshold of 50 scientists per discipline and were omitted from further analysis.

Table 3 provides a short description of variables used in the analysis ("Observatory" refers to the Polish Science Observatory and "Ministry" means the Polish Ministry of Education and Science).

**Table 3** Variables used in the analysis

| No | Variable | Description | Source |
|---|---|---|---|
| 1 | Biological age | Numerical variable. Biological age as provided by the national registry of scientists ($N=99{,}935$). Age in full years as of 2017 is used | Observatory |
| 2 | Age group | Categorical variable. Three age groups are used: young (39 and younger; $N=8400$), middle-aged (40–54; $N=11{,}014$), and older (55 and older; $N=6049$) scientists | Observatory |
| 3 | Gender | Binary variable, male or female, as provided by the national registry of scientists ($N=99{,}935$). No other options are possible in the registry | Observatory |
| 4 | Discipline | Categorical variable. All scientists ascribed to one of 27 Scopus ASJC (All Science Journal Classification) disciplines. Dominant disciplines were used ($N=25{,}463$) | Scopus |
| 5 | STEM disciplines | Categorical variable. STEM disciplines: AGRI, agricultural and biological sciences; BIO, biochemistry, genetics, and molecular biology; CHEMENG, chemical engineering; CHEM, chemistry; COMP, computer science; DEC, decision science; EARTH, earth and planetary sciences; ENER, energy; ENG, engineering; ENVIR, environmental science; IMMU, immunology and microbiology; MATER, materials science; MATH, mathematics; NEURO, neuroscience; NURS, nursing; PHARM, pharmacology, toxicology, and pharmaceutics; and PHYS, physics and astronomy. NEURO and NURS were omitted from the analysis as they did not meet an arbitrary minimum threshold of 50 scientists per discipline | Scopus |
| 6 | Non-STEM disciplines | Categorical variable. Non-STEM disciplines: BUS, business, management, and accounting; DENT, dentistry; ECON, economics, econometrics, and finance; HEALTH, health professions; HUM, arts and humanities; MED, medicine; PSYCH, psychology; SOC, social sciences; and VET, veterinary | Scopus |
| 7 | Male- and female-dominated disciplines | Binary variable. Male-dominated disciplines are those in which the percentage of male scientists exceeds or equals 50% ($N=12{,}786$ scientists). Female-dominated disciplines are those in which the percentage of female scientists exceeds 50% ($N=12{,}677$ scientists) | Observatory |
| 8 | Mean publication prestige (percentile rank) | Numerical variable. Mean prestige represents the median prestige value for all publications written by a scientist in the study period of 2009–2018. For journals for which the Scopus database did not ascribe a percentile rank, we have ascribed the percentile rank of 0 | Scopus |
| 9 | Research-intensive institution | Binary variable. The 10 institutions are the IDUB (or "Excellence Initiative–Research University") institutions selected in 2019 | Ministry |

The key methodological step was to determine what we termed an "individual publication portfolio" for every internationally visible Polish scientist (for the decade of 2009–2018). Next, using an individual scientist as the unit of analysis, we calculated the proportion of solo articles among all articles within the individual publication portfolio of every Polish scientist in the sample.

Thus, for all scientists, male and female, we constructed what we termed the individual publishing solo rate (for scientists publishing all their articles alone, the rate is 1) as a numerical variable. Analogously, a rate of 0 is equivalent to conducting no solo research: the scientist collaborates with others in all publications, that is, there are only collaborative articles in the portfolio. We have also considered another operationalization of the gender solo research gap (as in studies of gender wage gap, see Blau & Kahn, 2000; Sitzmann & Campbell, 2021). We considered computing 1 minus women's publications as a percentage of men's publications, with higher scores indicating a larger gap (Sitzmann & Campbell, 2021). However, comparing male and female solo rates across the various dimensions was more intuitive than comparing scores across them mostly because publishing patterns cannot be easily compared to other discriminatory practices in universities (such as tenure gap or salary gap). We are examining gender differences in publishing patterns within a national system from a cross-sectional perspective and we do not compare changing rates over time and across countries for which changing scores would be useful.

For the vast majority of scientists (10,015 or 67.3% of all men and 7690 women or 72.7% of all women; 17,705 in total), the individual publishing solo rate was zero, meaning they had not had a solo article published in the decade studied. The total number of solo articles in the study period is 18,900, where 12,781 (or 67.6%) were published by men and 6119 (or 32.4%) by women. The total number of solo scientists is 7758, of which the majority are men (Table 4).

We have decided not to analyze the time trend of solo work in 2009–2018 by gender and by the dimensions we use for individual publication portfolios in "Solo research, academic discipline and institutional research intensity by gender " to "A modeling approach: a fractional logit regression model" Sections. However, we have calculated individual solo rate for each year separately and the trend is clearly away from publishing solo, with the individual solo rates of 0.098 in 2009 and 0.068 in 2018 (down from 0.074 to 0.057 for women and down from 0.113 to 0.076 for men in this period). Using the data for particular years 2009–2018 rather than for the whole period is less reliable because Polish scientists tend not to publish, and not publish solo research, every year: in 2009 there were 9326 scientists and in 2018 16,805 scientists with at least one publication, the number of scientists increasing every year. However, the number of scientists with at least a single solo article in a given year was only 1165 (836 men and 329 women) in 2009 and 1540 (1004 men and 536 women) in 2018, also increasing every year. In terms of disciplines, the solo rate clearly decreases over time: the share of solo publications in individual portfolios has been clearly declining between 2009 and 2018 for every discipline (see Table 11). Polish scientists publish similar numbers of solo publications per year but more team publications per year; consequently, the solo rate decreases over time in every discipline.

**Table 4** Distribution of Polish scientists by publication type (solo scientists, non-solo scientists) and gender. Solo scientists are scientists with at least one solo article in their individual publication portfolio in the period studied (2009–2018) ($N = 25{,}463$ scientists)

| | Female Scientists | | | Male Scientists | | | Total | | |
|---|---|---|---|---|---|---|---|---|---|
| | *N* | Row % | Col % | *N* | Row % | Col % | *N* | Row % | Col % |
| Solo scientists (with at least one solo article in individual publication portfolios) | 2887 | 37.2 | 27.3 | 4871 | 62.8 | 32.7 | 7758 | 100.0 | 30.5 |
| Non-solo scientists (with no solo article in individual publication portfolios) | 7690 | 43.4 | 72.7 | 10,015 | 56.6 | 67.3 | 17,705 | 100.0 | 69.5 |
| Total | 10,577 | 41.5 | 100.0 | 14,886 | 58.5 | 100.0 | 25,463 | 100.0 | 100.0 |

The major difference between approaching solo research via individual publication portfolios and via aggregated percentages of solo research in the Polish science system as a whole is the role of publishing outliers, or highly productive scientists (see Kwiek, 2016). Their role is reduced in the first method, whereas they may play an excessive role in the second method, leading to distortions, especially in view of previous research showing a highly skewed distribution of productivity in Polish science (in which 10% of scientists produce about 50% of publications; Kwiek, 2018b). In the first method, each scientist has a clearly defined individual publication portfolio, with a specific individual publishing solo rate ranging from 0 to 1. The impact on the average male and female rates in Poland of scientists with 100 publications equals the impact of those with 10 publications.

Because of the relatively large sample used in our analysis ($N=25{,}463$), the results of the statistical tests were strengthened by the effect size analysis. When using large samples, classical statistical procedures usually lead to the rejection of the null hypothesis, even when differences between subpopulations are negligible (Leech et al., 2015: 196): "statistical significance is not the same as practical significance or importance. With large samples, you can find statistical significance even when the differences or associations are very small/weak" the phenomenon which is termed effect size. This may lead to considering small deviations as practically significant, while for large samples, they do not have to be so. Therefore, we examined statistical significance and effect size, and based on the results of the analysis, the practical significance of gender differences was found. Effect size is defined as "the strength of the relationship between the independent variable and the dependent variable and/or the magnitude of the difference between levels of the independent variable with respect to the dependent variable" (Leech et al., 2015: 197, see Cohen, 1988).

Kolmogorov-Smirnow tests for solo rate distribution normality were used and the null hypotheses were rejected for the whole population, as well as for all subpopulations analyzed (academic disciplines, institutional types, and age groups). Statistical assumptions of normality distribution were not met in any case and therefore we decided to use nonparametric Mann–Whitney test. Because we used the Mann–Whitney test to assess the difference between solo rate for male and female scientists, an $r$ coefficient was used (not to be confused with Pearson's $r$) to assess practical significance of these differences. The formula used was as follows:

$$r = \frac{|Z|}{\sqrt{n_1 + n_2}}$$

where $Z$ is test statistics, $n_1$ is size of first subsample, $n_2$ is the size of the second subsample.

The coefficient has the following interpretation: $r<0.1$ no effect, $0.1\leq r<0.3$ small effect, $0.3\leq r<0.5$ medium effect, and $r\geq 0.5$ large effect (Field et al., 2012: 58).

## Sample

The sample ($N=25,463$) consists of 14,886 male scientists and 10,577 female scientists (58.5% and 41.5%, respectively). It contains all scientists full-time employed in the higher education sector who had at least a single article indexed in the Scopus database in the period of 2009–2018 and who had at least a doctoral degree. Thus, the sample includes all internationally visible (through publication type: article) Polish academic scientists. The steps in defining the sample were as follows: scientists from all public science sectors: 99,935; scientists with doctorates: 70,272; scientists with doctorates and employed in higher education: 54,448, of which scientists with publications of any type in Scopus in 2009–2018: 32,937; scientists with journal articles only: 25,463, which is our sample.

In terms of the age distribution, about half of them are middle-aged (or in the 40–54 age bracket; 49.7%), and in terms of academic positions, over half of them are assistant professors (56.0%). Table 10 shows column percentages, which enable the analysis of the gender distribution by major age groups, academic positions, and disciplines (by type: STEM and non-STEM, female-dominated and male-dominated), and it shows row percentages, which enable the analysis of how male and female scientists are distributed according to a given feature. About half the scientists work in female-dominated disciplines and about half in male-dominated disciplines (49.8% and 50.2%). All assistant professors hold doctoral degrees, all associate professors hold habilitations, and all full professors hold professorship titles.

## Limitations

Our research is affected by a selection bias as a result of the database construction: we select only internationally visible scientists from higher education institutions, that is, scientists with Scopus-indexed articles, and we select only scientists with at least doctorates. So our scientists would come from larger academic centers where international publications are required rather than from smaller institutions where they are not obligatory in academic careers (except for the humanities where they are generally not required across the system). Doctorates are generally required to become part of the academic profession in Poland. There are also five simplifying assumptions (as in Kwiek & Roszka, 2021b), which are as follows: (1) We examine a decade of individual publishing output, although the actual publishing period may be shorter for younger scientists; (2) Scopus-provided journal percentile ranks are deemed stable, although they may fluctuate over the period studied; (3) we assume that scientists did not change institutions in the decade studied; (4) we regard scientists who were assistant, associate, and full professors on the date of reference (November 21, 2017) as keeping these positions for the decade studied, while these positions are the highest ranks achieved in the study period; and (5) for the purpose of international comparability in the results, we refer to three categories of academic positions (assistant, associate, and full professor), although in practice, two Polish academic degrees (doctorate and habilitation) and a Polish academic title (professorship) are used. Thus, academic positions are useful proxies for Polish academic degrees and titles. While the administrative and biographical variables of biological age, academic position, employment type, and institution were defined as of November 21, 2017, the publication and citation variables derived from the Scopus database were constructed to show mean values for the decade of 2009–2018 (and they may have differed from year to year). Therefore, another limitation is that the values for 2017 for some variables and the mean values for the decade of 2009–2018 are used in the same analysis.

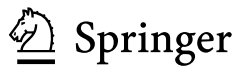

## Results

### Solo research, academic discipline and institutional research intensity by gender

Based on previous literature, we test two hypotheses:

**H1** *Academic disciplines* We expect that female scientists will exhibit lower individual publishing solo rates than male scientists across all academic disciplines.

**H2** *Institutional research intensity* We expect that female scientists will exhibit higher individual publishing solo rates than male scientists in research-intensive institutions and lower individual publishing solo rates than male scientists in institutions less involved in research.

We use two approaches to individual publishing solo rates: we examine rates by gender for all scientists and for solo scientists only. The individual publishing solo rate is the fraction of solo articles among all articles within the individual publication portfolio for a decade of 2009–2018, ranging from 0 to 1 (for no solo articles and all solo articles in individual publication portfolios, respectively). Solo scientists are scientists with at least one solo article in their individual publication portfolio in the period studied (2009–2018); therefore their individual solo publishing rate is higher than 0.

The former approach highlights the distribution of the rate among all scientists by gender and discipline (left panel in Table 5), the rate being as low as 0.013–0.016 in BIO (or less than 2% of single-authored publications in an individual publication portfolio) and as high as 0.50–0.53 in SOC and 0.76 in HUM (or as much as 50–76%), depending on gender. The latter approach highlights the intensity of publishing solo, or the individual publishing solo rate only for those scientists who actually ever published solo, by discipline (right panel in Table 5).

To take a generally high-collaboration discipline of chemistry (CHEM) and a generally low-collaboration discipline of arts and humanities (HUM), it can be observed that, in the former, 3.2–3.4% of publications in the individual publication portfolios of male and female scientists are published solo; in the latter, the percentage is about 76% (Fig. 2, left panel)). However, for solo scientists only, for CHEM, the intensity is 15–16%, and for HUM, it is 92–93% (right panel). In CHEM, solo scientists publish solo occasionally, compared with HUM, where solo scientists publish almost all their articles solo. In other words, in HUM, three-quarters of scientists publish solo, and for those who ever publish solo (in the decade studied), the pattern of solo publishing is intense: more than 90% of their articles are solo articles.

In purely descriptive terms, female scientists across disciplines publish solo only slightly less often than men do; however, when they do, they publish solo with higher intensity in both heavily male-dominated disciplines (e.g., physics and astronomy (PHYS) with 16.6% of women publishing solo; see the gender distribution of scientists across 14 male-dominated and 10 female-dominated disciplines in Table 10; earth and planetary sciences (EARTH), with 33.4% of women publishing solo) and female-dominated disciplines (e.g., pharmacology, toxicology, and pharmaceutics (PHARM), with 66.5% of women publishing solo; medicine (MED), with 53.7% of women publishing solo). Additionally, 67.3% of males and 72.7% of women had not had a solo article published in the decade studied (their individual publishing solo rate was zero).

**Table 5** Mean individual publishing solo rate (range: 0–1) by discipline and gender: for all scientists in the sample (left panel) and for solo scientists only. Solo scientists are scientists with at least one solo article in their individual publication portfolio in the period studied (2009–2018) ($N = 25,463$ scientists)

| All scientists | | | | | | | Scientists with at least one solo article (solo scientists) | | | | | | |
|---|---|---|---|---|---|---|---|---|---|---|---|---|---|
| Discipline | Male | Female | Total | $Z$ | $p$ | $r$ | Discipline | Male | Female | Total | $Z$ | $p$ | $r$ |
| HUM | 0.758 | 0.760 | 0.759 | −0.233 | 0.816 | 0.007 | HUM | 0.919 | 0.932 | 0.925 | −0.93 | 0.352 | 0.032 |
| SOC | 0.529 | 0.496 | 0.513 | −1.14 | 0.254 | 0.036 | SOC | 0.787 | 0.764 | 0.775 | −0.956 | 0.339 | 0.037 |
| DEC | 0.430 | 0.326 | 0.384 | −1.016 | 0.31 | 0.138 | DEC | 0.716 | 0.870 | 0.768 | −1.506 | 0.132 | 0.290 |
| ECON | 0.364 | 0.358 | 0.361 | −0.001 | 0.999 | 0.000 | HEALTH | 0.581 | 0.820 | 0.690 | −1.344 | 0.179 | 0.405 |
| MATH | 0.279 | 0.248 | 0.271 | −1.468 | 0.142 | 0.046 | ECON | 0.688 | 0.659 | 0.673 | −0.649 | 0.516 | 0.046 |
| BUS | 0.278 | 0.255 | 0.266 | −0.425 | 0.671 | 0.016 | BUS | 0.679 | 0.625 | 0.651 | −1.427 | 0.153 | 0.084 |
| PSYCH | 0.180 | 0.168 | 0.173 | −0.026 | 0.979 | 0.001 | ENER | 0.482 | 0.611 | 0.513 | −1.645 | 0.1 | 0.173 |
| EARTH | 0.160 | 0.178 | 0.166 | −0.397 | 0.691 | 0.012 | PSYCH | 0.471 | 0.453 | 0.460 | −0.579 | 0.563 | 0.054 |
| ENER | 0.154 | 0.164 | 0.157 | −0.509 | 0.611 | 0.030 | MATH | 0.441 | 0.429 | 0.438 | −0.372 | 0.71 | 0.015 |
| COMP | 0.157 | 0.153 | 0.156 | −0.258 | 0.796 | 0.008 | COMP | 0.428 | 0.433 | 0.429 | −0.244 | 0.807 | 0.013 |
| ENG | 0.150 | 0.165 | 0.152 | −1.068 | 0.286 | 0.018 | EARTH | 0.377 | 0.431 | 0.395 | −2.437 | 0.015 | 0.111 |
| HEALTH | 0.079 | 0.178 | 0.113 | −0.983 | 0.326 | 0.120 | ENG | 0.377 | 0.395 | 0.380 | −1.056 | 0.291 | 0.029 |
| CHEMENG | 0.099 | 0.047 | 0.079 | −2.175 | 0.03 | 0.099 | CHEMENG | 0.375 | 0.247 | 0.335 | −2.252 | 0.024 | 0.212 |
| ENVIR | 0.071 | 0.080 | 0.076 | −2.783 | 0.005 | 0.068 | IMMU | 0.122 | 0.315 | 0.267 | −0.466 | 0.642 | 0.135 |
| PHYS | 0.071 | 0.067 | 0.070 | −0.709 | 0.478 | 0.021 | ENVIR | 0.281 | 0.251 | 0.264 | −1.044 | 0.296 | 0.048 |
| MATER | 0.050 | 0.047 | 0.049 | −0.036 | 0.971 | 0.001 | MED | 0.191 | 0.283 | 0.239 | −3.516 | <0.001 | 0.198 |
| AGRI | 0.046 | 0.043 | 0.044 | −0.478 | 0.633 | 0.009 | AGRI | 0.237 | 0.234 | 0.236 | −0.705 | 0.481 | 0.031 |
| CHEM | 0.032 | 0.034 | 0.033 | −1.298 | 0.194 | 0.034 | BIO | 0.264 | 0.206 | 0.228 | −0.883 | 0.377 | 0.083 |
| IMMU | 0.013 | 0.032 | 0.027 | −0.024 | 0.981 | 0.002 | PHYS | 0.213 | 0.230 | 0.216 | −1.689 | 0.091 | 0.089 |
| Med | 0.017 | 0.024 | 0.021 | −0.486 | 0.627 | 0.008 | MATER | 0.190 | 0.176 | 0.185 | −0.475 | 0.635 | 0.024 |
| PHARM | 0.009 | 0.019 | 0.016 | −0.147 | 0.883 | 0.009 | PHARM | 0.089 | 0.200 | 0.160 | −2.18 | 0.029 | 0.436 |
| VET | 0.011 | 0.020 | 0.015 | −0.804 | 0.421 | 0.044 | CHEM | 0.163 | 0.154 | 0.158 | −1.948 | 0.051 | 0.111 |
| BIO | 0.016 | 0.013 | 0.014 | −0.325 | 0.745 | 0.008 | VET | 0.100 | 0.241 | 0.151 | −2.041 | 0.041 | 0.355 |
| DENT | 0.005 | 0.005 | 0.005 | −0.032 | 0.975 | 0.004 | DENT | 0.091 | 0.100 | 0.097 | −0.447 | 0.655 | 0.224 |

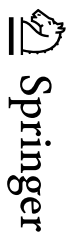

**Table 5** (continued)

| All scientists | | | | | | | Scientists with at least one solo article (solo scientists) | | | | | | |
|---|---|---|---|---|---|---|---|---|---|---|---|---|---|
| Discipline | Male | Female | Total | *Z* | *p* | *r* | Discipline | Male | Female | Total | *Z* | *p* | *r* |
| Total | 0.143 | 0.130 | 0.138 | − 8.227 | < 0.001 | 0.052 | Total | 0.437 | 0.477 | 0.452 | − 3.968 | < 0.001 | 0.045 |

Male scientists / Female scientists

| ASJC | All: Female scientists | All: Male scientists | Solo only: Female scientists | Solo only: Male scientists |
|---|---|---|---|---|
| HUM | 0.76 | 0.76 | 0.93 | 0.92 |
| SOC | 0.50 | 0.53 | 0.76 | 0.79 |
| DEC | 0.33 | 0.43 | 0.87 | 0.72 |
| ECON | 0.36 | 0.36 | 0.66 | 0.69 |
| BUS | 0.26 | 0.28 | 0.62 | 0.68 |
| HEALTH | 0.18 | 0.08 | 0.82 | 0.58 |
| MATH | 0.25 | 0.28 | 0.43 | 0.44 |
| ENER | 0.16 | 0.15 | 0.61 | 0.48 |
| PSYCH | 0.17 | 0.18 | 0.45 | 0.47 |
| COMP | 0.15 | 0.16 | 0.43 | 0.43 |
| EARTH | 0.18 | 0.16 | 0.43 | 0.38 |
| ENG | 0.16 | 0.15 | 0.40 | 0.38 |
| CHEMENG | 0.05 | 0.10 | 0.25 | 0.38 |
| ENVIR | 0.08 | 0.07 | 0.25 | 0.28 |
| IMMU | 0.03 | 0.01 | 0.31 | 0.12 |
| PHYS | 0.07 | 0.07 | 0.23 | 0.21 |
| AGRI | 0.04 | 0.05 | 0.23 | 0.24 |
| MED | 0.02 | 0.02 | 0.28 | 0.19 |
| BIO | 0.01 | 0.02 | 0.21 | 0.26 |
| MATER | 0.05 | 0.05 | 0.18 | 0.19 |
| CHEM | 0.03 | 0.03 | 0.15 | 0.16 |
| PHARM | 0.02 | 0.01 | 0.20 | 0.09 |
| VET | 0.02 | 0.01 | 0.24 | 0.10 |
| DENT | 0.01 | 0.01 | 0.10 | 0.09 |

Mean solo rate

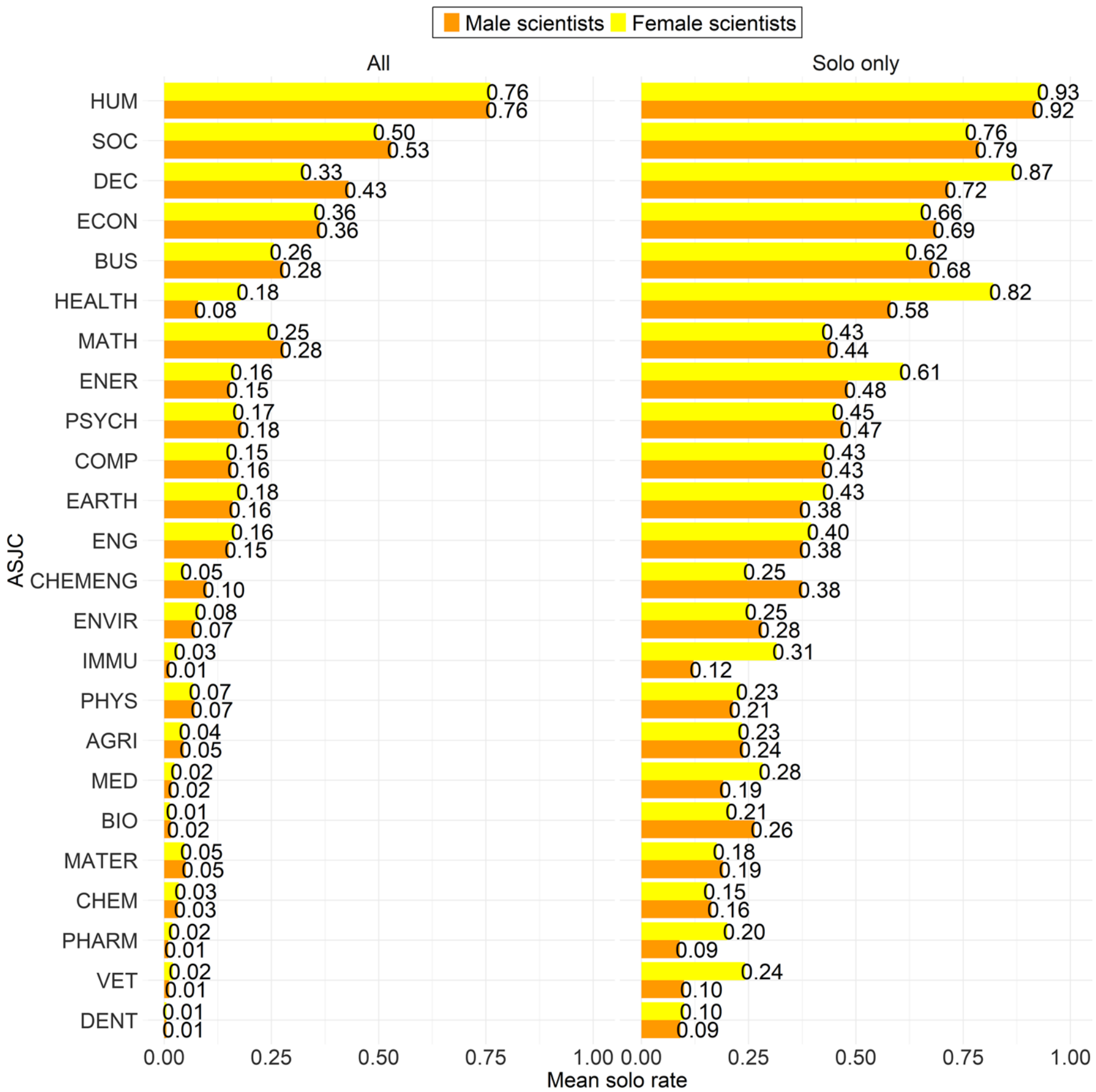

**Fig. 2** Mean individual publishing solo rate (range: 0–1) for all scientists in the sample (left panel) and for solo scientists only (right panel) by gender and discipline (in descending order). Solo scientists are scientists with at least one solo article in their individual publication portfolio in the period studied (2009–2018) ($N=25{,}463$ scientists)

However, overall, the differences within disciplines by gender are smaller than expected and statistically insignificant, except for several cases. Interestingly, both for all scientists combined and for solo scientists combined (Total), differences between men and women in solo rates are statistically significant; and for the disciplines with statistically significant differences between men and women (CHEMENG and ENVIR for all scientists, and CHEMENG, MED, PHARM, CHEM and VET for solo scientists), the differences are also significant.

At a closer look, the differences are statistically significant but with no practical significance (in terms of the effect-size *r* coefficient, see the last columns in both panels). As Leech et al., (2015: 196) put it, “a statistically significant result with a small effect size means that we can be very confident that there is *some* difference or association, but it is probably small and may not be practically important”. For all scientists, *r* for CHEMENG does not exceed 0.1 which is the boundary value for observing any effect.

For solo scientists, the effect size is slightly larger (small) and exceeds 0.1 for as many as eleven domains, with statistically significant differences observed only for PHARM, VET, CHEMENG, MED, and EARTH. A medium effect size was observed only for PHARM and VET (due to relatively small sample sizes), while for large domains such as CHEMENG, MED, and EARTH the effect strength was weak.

In conclusion, while some differences in the propensity to write solo by discipline and gender are statistically significant, they are marginally important ("practically insignificant" in Leeches et al.'s terms) for virtually all disciplines. The individual publication portfolios of men and women are remarkably similar, even allowing for their differences in the fraction of publishing solo. The results of our complementary analysis of the *r* coefficient to assess the effect size of the association between solo rate for male and female scientists shows that no statistical difference exists between them.

Previous literature indicates differences in publishing solo by institutional type: the more research focused an institution, the higher the involvement in publishing solo among faculty (e.g., Vafeas, 2010: 340). Therefore, we test whether the individual solo rate differs by gender and institutional type. We contrast ten IDUB research-intensive institutions with 75 other research-involved institutions. The ten IDUB institutions are the "Excellence Initiative–Research University" institutions, which were selected for additional research funding for the 2020–2026 period. The IDUB institutions include both top Polish universities and technical universities, and they were the top 10 Polish institutions in terms of total publications output in 2009–2018 (articles only).

We compare two classes of men and women in two institutional types: all scientists (upper panel) and solo scientists only (lower panel, Table 6). In terms of institutional research intensity, gender differences in solo rate proved statistically significant (except for Rest institutions for solo scientists only). However, an analysis of solo rate for all scientists shows that the differences are small (0.38 pp. for IDUB institutions and 1.77 pp. for Rest institutions); in this case, the statistical significance of the differences should be attributed to large numbers of observations in both subsamples. Indeed, the effect size analysis clearly indicates that the differences, although statistically significant, are not practically significant (the last column: $r = 0.043$; for $r < 0.1$, the coefficient interpretation is that there is no effect).

It is the same case with solo scientists only (lower panel): a relatively high gender difference in solo rate for IDUB solo scientists only (scientists with at least a single solo article in their publication portfolio and working in research-intensive institutions) reaches almost 10 pp. But the size of the effect is small ($r = 0.117$) despite statistical significance ($p < 0.001$). While the solo rate for all scientists for all institutions combined and for the rest of institutions is higher for men (Total and Rest, Table 6, upper panel), for research-intensive IDUB institutions, it is higher for women, in accordance with previous literature. However, the size of the effect is again small, as the effect size analysis indicates. The intensity of solo publishing (shown through the rate for solo scientists only) is substantially higher for women in IDUB research-intensive institutions and for women in all institutions combined. Even though the differences are statistically significant, the effect size is weak.

**Table 6** Mean individual publishing solo rate (range: 0–1) by institutional type and gender: for all scientists in the sample (top panel) and solo scientists only (bottom panel). Solo scientists are scientists with at least one solo article in their individual publication portfolio in the period studied (2009–2018). The Mann–Whitney test was used ($N = 25{,}463$ scientists)

| | Institutional type | Female | Male | $Z$ | $p$ | $r$ |
|---|---|---|---|---|---|---|
| All scientists | IDUB: research-intensive | 0.1540 | 0.1502 | 3.920 | < 0.001 | 0.043 |
| | Rest | 0.1211 | 0.1388 | 6.311 | < 0.001 | 0.048 |
| | Total | 0.1301 | 0.1429 | 8.277 | < 0.001 | 0.052 |
| Solo scientists only | IDUB: research-intensive | 0.5030 | 0.4093 | 6.262 | < 0.001 | 0.117 |
| | Rest | 0.4651 | 0.4549 | 0.157 | 0.875 | 0.002 |
| | Total | 0.4767 | 0.4366 | 3.968 | < 0.001 | 0.045 |

Neither Hypothesis 1 on differences in the individual solo rate by gender and academic disciplines nor Hypothesis 2 on differences in the rate by gender and institutional research intensity found support in our data: although some differences between male and female scientists are statistically significant, they have none or small practical significance.

## Solo research, biological age, and gender

Finally, as part of our two-dimensional analyses, based on previous literature, we test the following hypothesis:

**H3** *Biological age* We expect that younger male and female scientists will exhibit higher individual publishing solo rates than older male and female scientists do.

There is a unique variable available for each observation in our study: biological age (strongly correlated with academic positions, not explored in this paper). We now examine the individual publishing solo rate by gender and (1) age groups and (2) age in two versions, as above for all scientists and for solo scientists only. We divided our sample into the following three age categories: young scientists (under 40), middle-aged scientists (aged 40–54), and older scientists (aged 55 and older); of these, middle-aged scientists form the largest age group (45.79%). The proportion of men and women is almost equal among young scientists but women comprise less than 30% of older scientists (see Table 10).

The gender differences in solo publishing patterns by age group are as follows (Table 7): for all scientists (left panel), the rate for young male scientists is higher than for young female scientists: they are more involved in solo publishing. The highest level of solo publishing is noted for middle-aged scientists, for both sexes. Younger scientists have significantly lower individual solo publishing rates than middle-aged scientists, and the differences are higher for women than for men. However, female scientists in all age groups are more intensely involved in solo publishing (right panel): female solo scientists show considerably higher rates than male solo scientists do. For instance, for young scientists, female solo scientists have a rate of 0.485, whereas that for male solo scientists is 0.434; that is, for female scientists who have ever solo-authored, 48.5% of publications in their individual publication portfolios are published solo compared with 43.4% for male scientists. For older scientists, the difference is 51.02% versus 44.73%.

However, while all above differences are statistically significant, we test whether the differences also have practical significance in terms of the effect size $r$ coefficient. The analysis shows that $r$ indicates no effect for all gender differences and small effect in one case only: young academics.

A year-by-year approach illustrated by regression lines generally confirms the two similar trends for both genders (Fig. 3). For all scientists, the generally upward trend in the individual solo rate between 0.05 and 0.15 for male scientists lasts until the age of 40 and for female scientists lasts until the age of 55 (see lower lines in both panels). For both genders, the rate drops for scientists between 60 and 70, in a similar manner. However,

**Table 7** Mean individual publishing solo rate (range: 0–1) for all scientists in the sample (left panels) and for solo scientists only (right panels), by gender and age group. ($N=25{,}463$ scientists)

| Age group | All scientists | | | | | | Scientists with at least one solo article (solo scientists) | | | | | |
|---|---|---|---|---|---|---|---|---|---|---|---|---|
| | Male | Female | Total | *Z* | *p* | *r* | Male | Female | Total | *Z* | *p* | *r* |
| Young ($<40$) | 0.132 | 0.104 | 0.118 | −8.152 | $<0.001$ | 0.102 | 0.434 | 0.485 | 0.455 | −2.760 | 0.006 | 0.070 |
| Middle-aged (40–54) | 0.158 | 0.146 | 0.153 | −4.962 | $<0.001$ | 0.044 | 0.432 | 0.463 | 0.444 | −2.124 | 0.034 | 0.032 |
| Older (55 and more) | 0.130 | 0.136 | 0.132 | −1.457 | 0.145 | 0.018 | 0.447 | 0.510 | 0.464 | −2.645 | 0.008 | 0.062 |
| Total | 0.143 | 0.130 | 0.138 | −8.227 | $<0.001$ | 0.052 | 0.437 | 0.477 | 0.452 | −3.968 | $<0.001$ | 0.045 |

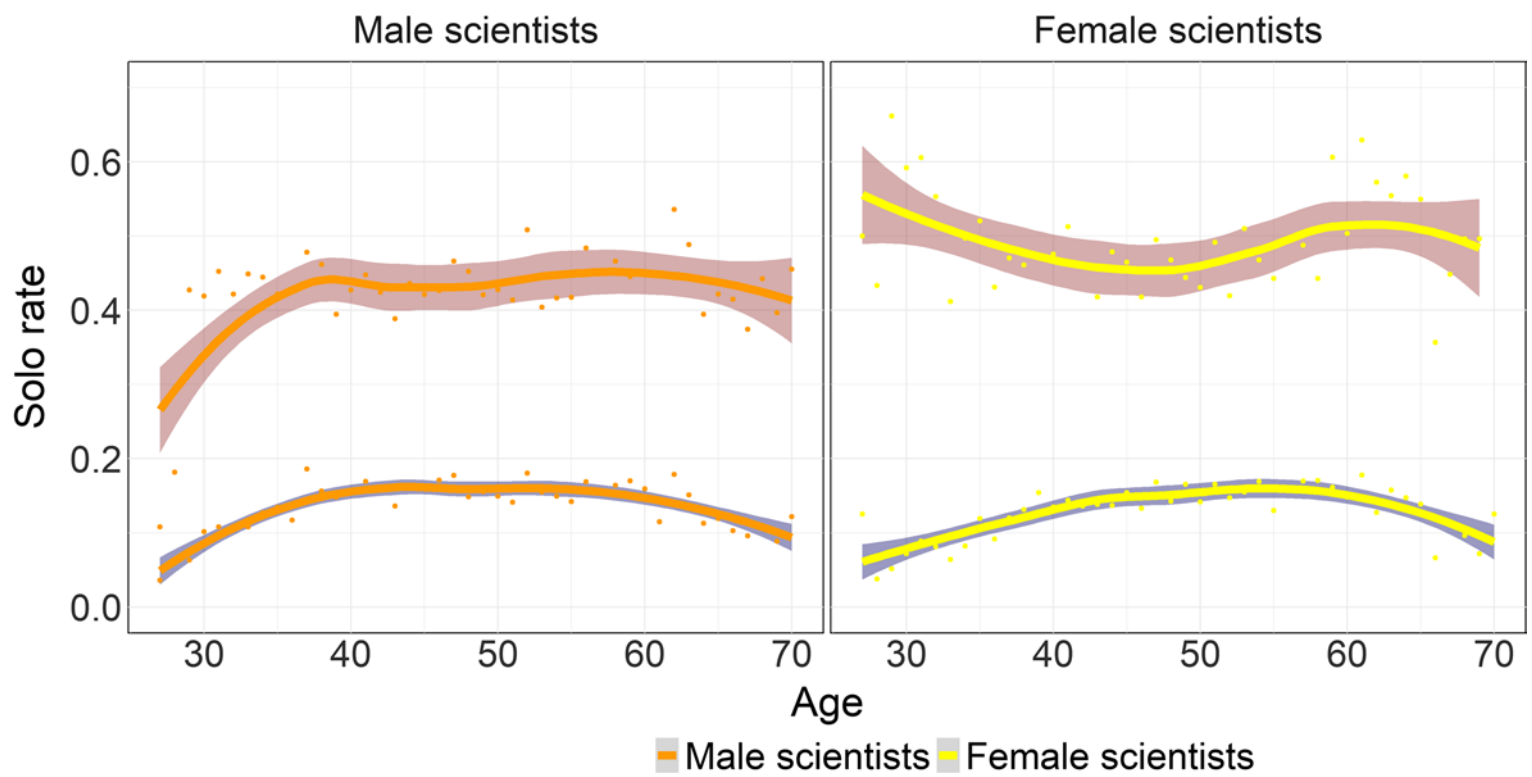

**Fig. 3** Mean individual publishing solo rate (range 0–1) by gender and age for all scientists in the sample (lower lines) and for solo scientists only (upper lines). The regression line was estimated using the method of local polynomial regression fitting. The gray area represents 95% confidence intervals. Each year of age is represented by a single dot (a cut-off point of 70 is used). Dots represent mean values ($N = 25{,}463$ scientists)

the intensity of solo publishing (i.e., the rate for solo scientists only) for female scientists is equal or higher for each age (see higher lines in both panels); specifically, it is much higher for young scientists in their 30s. In a specific Polish case, scientists of this age have just received their doctoral degrees. Solo female scientists in their 30s have a substantially higher share of solo articles in their publication portfolios, the highest difference for women being in their early 30s. Then, in their 40s, the gender differential in solo publishing intensity is marginal, increasing again for women in their 50s.

The most notable gender differential in solo publishing intensity is for scientists in their 30s and 60s when the rate is higher, in the beginning and at the end of academic careers. The dots in Fig. 3 represent the median value of the solo rate for each year of age. Our Hypothesis 3 about gender and biological age did not find support in our data: gender differences in the solo rate turn out to have no practical significance.

## A modeling approach: a fractional logit regression model

In our multi-dimensional approach, our hypothesis is as follows:

> *H4: Solo publication propensity* We expect that being a female scientist decreases the propensity to conduct solo research or the individual publishing solo rate (in fractional logit regression models).

Finally, we use a regression model for a fractional dependent variable: a fractional logit regression model (Papke and Woolridge, 1996), designed for variables bounded between 0 and 1 (as with our dependent variable: the solo research ratio). The standard practice of using linear models to examine how a set of explanatory variables influences a given proportion or fractional response variable is not appropriate here (Ramalho et al., 2011: 19). In this model, no special data adjustments are needed for the extreme values of 0 and 1. As our dependent variable is fractional (ranging from 0 to 1), we estimate a fractional logit regression model. We estimate odds ratios for conducting solo research, that is, publishing solo articles. We calculate the solo rate as the percentage of solo articles in all the published articles in all the scientists' individual publication portfolios. Using partial effects of fractional logistic regression approach, we estimated the probability of conducting solo research.

In the model, we use both individual-level and organizational-level predictors. Individual-level predictors are gender, age, academic position (expressed through the proxies of doctorate, habilitation, and professorship), dominant ASJC discipline (STEM or non-STEM), average journal prestige rate in a scientist's individual publication portfolio (range, 0–99), average individual productivity in the study period (average number of articles per year, full counting method used), international collaboration rate (in the individual publication portfolio), average team size (mean value of number of collaborators per article in all articles from the study period), and publishing in a male-dominated discipline (male-dominated or female-dominated). The only organizational-level predictor used in the models is highly research-intensive (10 IDUB higher education institutions and the remaining 75 institutions).

We have estimated four models for four distinct populations for all academic ranks ($N=24{,}467$) and separately for the three ranks (for full professors, $N=3508$; associate professors, $N=7122$; and for assistant professors, $N=13{,}837$). The selected predictors of the individual publishing solo rate in the estimated models explain a very high percentage of the variability of the dependent variable, from 77.5% in the model for associate professors to 82.3% in the model describing relationships in the population of full professors (Table 8). At the same time, it is worth noting that gender does not explain the rate's variability in any of the models (at the significance level $\alpha=0.05$). The strongest predictor for each population studied was the average team size. An increase in the value of this variable by one author resulted in an average decrease in the rate by 8–11 percentage points (pp) (all other things held constant), depending on the model, which is an order of magnitude higher than other predicators. In addition, in each analyzed population, publishing in STEM fields negatively affects the propensity to publish solo by 3–4 pp on average. The occurrence of collinearity was checked by analyzing the values lying on the main diagonal of the inverted correlation matrix of independent variables. The empirical range of variability of these values ranged from 1 to 2 (see the variance inflation factor, the VIF column in Table 8: VIF provides an index that measures how much the variance of an estimated regression coefficient is increased because of collinearity), which indicates a negligible correlation of independent features.

Although gender is not a significant predictor of the rate, publishing in (quantitatively) male-dominated disciplines has a significant and positive impact on the variable, explained by 1–3 percentage points. An inverse relationship of similar strength can be observed in the influence of the international collaboration rate an increase of this variable by 1 unit results in an average individual publishing solo rate decrease of 1–4 pp, although it should be noted that, in the case of assistant professors, the influence of this predictor is not significantly different from zero. The influence of similar strength also occurs in the case of working in research-intensive institutions this results in an average individual publishing solo rate increase of slightly more than 1 pp, but in the case of full professors, the influence of this predictor is not statistically significant.

The average number of articles within a decade in the case of all populations has a significantly positive influence. An increase in the average number of articles by 1 causes an average rate increase of only 0.3–0.5 pp. Large productivity, however, can have a significant impact on the rate, as writing 100 articles in a decade results in an average increase of 30–50 pp. Definitely the weakest impact can be observed for age and average prestige. Both variables have a negative impact on the rate. Increase of average prestige by 1 unit causes average (in each population significant) rate decrease by 0.03 (assistant professors) to 0.09 (associate professors). In the case of age, this decrease is slightly smaller, from 0.02 (all scientists) to 0.15 (associate professors). The exception is assistant professors, for whom age is a positive predictor and causes an average rate increase of 0.07 pp with each completed year of life.

**Table 8** Four models

| | Estimate | SE | *t* Value | Pr(>\|*t*\|) | VIF |
|---|---|---|---|---|---|
| *Model 1. Scientists: All Ranks, N = 24,467. $R^2$: 0.786* | | | | | |
| Male | 0.0009 | 0.0019 | 0.441 | 0.659 | 1.131 |
| Age | −0.0002 | 0.0001 | −2.037 | 0.042 | 2.086 |
| Average prestige | −0.0005 | 0.0001 | −7.849 | <0.001 | 1.245 |
| Productivity in the study period | 0.0048 | 0.0008 | 5.937 | <0.001 | 1.206 |
| International collaboration rate | −0.0143 | 0.0052 | −2.753 | 0.006 | 1.208 |
| Publishing in a male-dominated discipl | 0.0195 | 0.0029 | 6.678 | <0.001 | 1.242 |
| Average team size | −0.1061 | 0.0021 | −49.485 | <0.001 | 1.294 |
| Full professor | 0.0007 | 0.0037 | 0.196 | 0.844 | 2.121 |
| Associate professor | 0.0256 | 0.0022 | 11.725 | <0.001 | 1.403 |
| IDUB | 0.0121 | 0.0018 | 6.591 | <0.001 | 1.053 |
| STEM | −0.0357 | 0.0029 | −12.310 | <0.001 | 1.120 |
| *Model 2. Full Professors, N = 3508. $R^2$: 0.823* | | | | | |
| Male | 0.0040 | 0.0055 | 0.730 | 0.466 | 1.077 |
| Age | −0.0009 | 0.0003 | −3.231 | 0.001 | 1.092 |
| Average prestige | −0.0007 | 0.0002 | −4.690 | <0.001 | 1.292 |
| Productivity in the study period | 0.0036 | 0.0015 | 2.320 | 0.020 | 1.207 |
| International collaboration rate | −0.0261 | 0.0107 | −2.451 | 0.014 | 1.290 |
| Publishing in a male-dominated discipl | 0.0234 | 0.0072 | 3.241 | 0.001 | 1.254 |
| Average team size | −0.0807 | 0.0047 | −17.289 | <0.001 | 1.266 |
| IDUB | 0.0050 | 0.0040 | 1.250 | 0.211 | 1.075 |
| STEM | −0.0457 | 0.0066 | −6.918 | <0.001 | 1.190 |
| *Model 3. Associate Professors, N = 7122. $R^2$: 0.775* | | | | | |
| Male | −0.0072 | 0.0039 | −1.831 | 0.067 | 1.082 |
| Age | −0.0015 | 0.0002 | −6.803 | <0.001 | 1.121 |
| Average prestige | −0.0009 | 0.0001 | −6.294 | <0.001 | 1.282 |
| Productivity in the study period | 0.0062 | 0.0024 | 2.519 | 0.012 | 1.252 |
| International collaboration rate | −0.0422 | 0.0111 | −3.813 | <0.001 | 1.201 |
| Publishing in a male-dominated discipl | 0.0359 | 0.0063 | 5.724 | <0.001 | 1.243 |
| Average team size | −0.1125 | 0.0050 | −22.447 | <0.001 | 1.325 |
| IDUB | 0.0116 | 0.0037 | 3.086 | 0.002 | 1.053 |
| STEM | −0.0409 | 0.0058 | −7.112 | <0.001 | 1.127 |
| *Model 4. Assistant Professors, N = 13,837. $R^2$: 0.791* | | | | | |
| Male | 0.0039 | 0.0024 | 1.639 | 0.101 | 1.119 |
| Age | 0.0007 | 0.0001 | 4.723 | <0.001 | 1.150 |
| Average prestige | −0.0003 | 0.0001 | −3.998 | <0.001 | 1.221 |
| Productivity in the study period | 0.0053 | 0.0011 | 5.013 | <0.001 | 1.186 |
| International collaboration rate | 0.0023 | 0.0068 | 0.338 | 0.735 | 1.193 |
| Publishing in a male-dominated discipl | 0.0130 | 0.0035 | 3.701 | <0.001 | 1.261 |
| Average team size | −0.1082 | 0.0025 | −43.772 | <0.001 | 1.299 |
| IDUB | 0.0130 | 0.0024 | 5.529 | <0.001 | 1.050 |
| STEM | −0.0287 | 0.0036 | −8.041 | <0.001 | 1.114 |

The interval estimation of model parameters indicates overlapping of estimation values for all variables and all models except the team size variable from the model for full professors. In this model, the size of the team has a significantly stronger impact than in the other models (Fig. 4). This means that the academic position does not significantly affect the rate. This is indicated by the estimates of Model 1 (for all scientists), where the position of full professor does not affect the rate at all, while for associate professor, although its influence is significantly different from zero, its strength is relatively small. In the case of age, it plays a positive (although weak) role for assistant professors, which can be explained by the willingness to gain independent output enabling promotion. A certain difference can also be mentioned in Model 3 (for associate professors) for the average international collaboration variable, where the position of associate professor is characterized by a significantly weaker influence on the rate than in Models 1 and 2 but almost entirely overlaps with the interval estimate for Model 4.

The analysis of residual components of the models shows that their distribution does not follow the normal distribution (see *K–S* test results in Table 9). However, distributions of the residuals are characterized by relatively small variability (they are strongly concentrated at zero value, see kurtosis values); there are numerous extreme values among them, but they do not significantly influence the distributions of residuals since the skewness values are close to 0. The number of values exceeding the extremes that is, not belonging to the range of $<-3,3>$ (which in the analysis of the residuals of regression models mean typical values based on the three-sigmas rule) is relatively small and oscillates between 1.23 (for assistant professors) and 2.65% (for full professors; see the outliers percentage in Table 9).

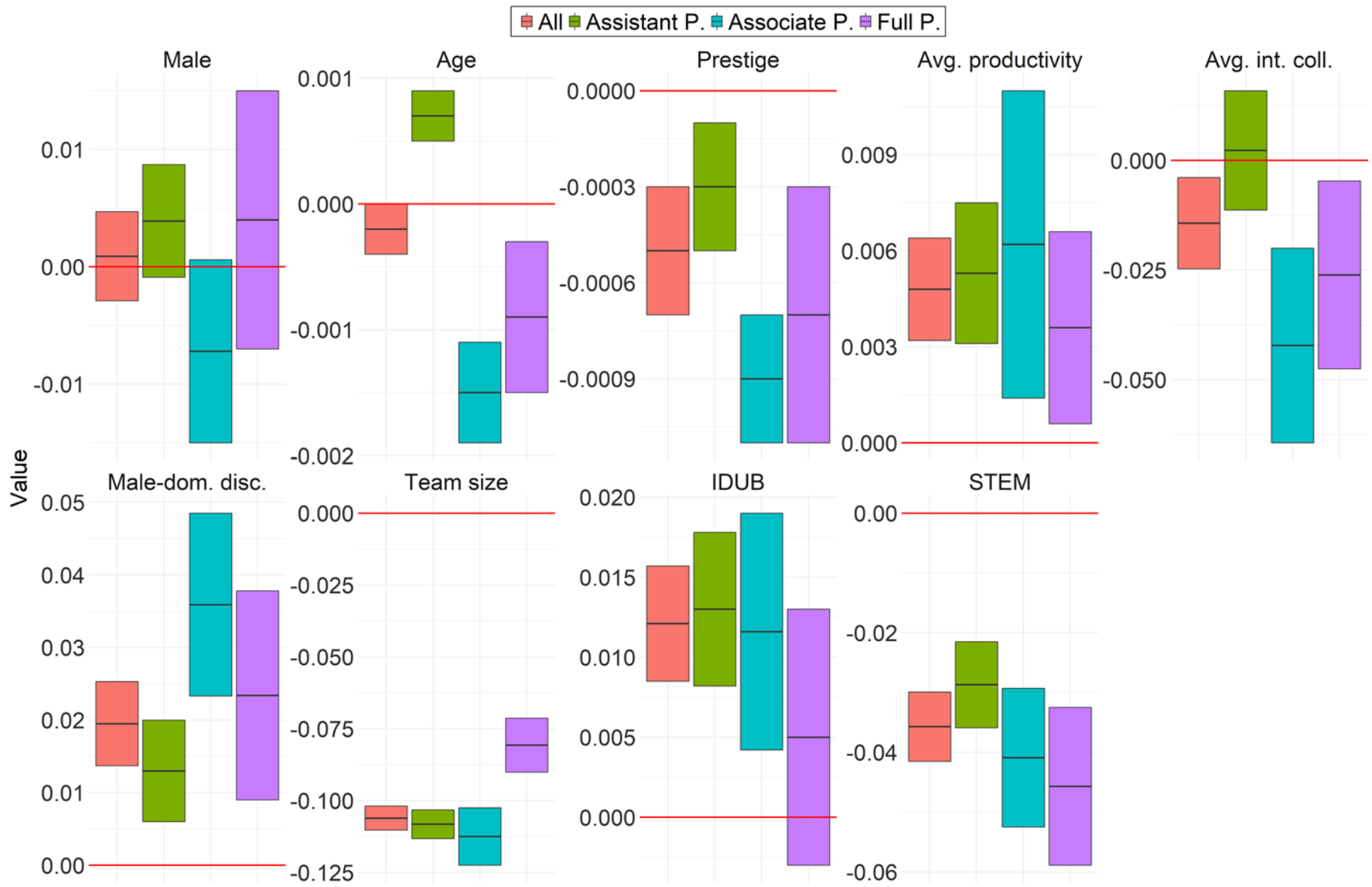

**Fig. 4** Confidence intervals range for models' parameters comparison of four models, all variables

**Table 9** Residuals statistics

| | All | Assistant Professor | Associate Professor | Full Professor |
|---|---|---|---|---|
| Minimum | −4.291 | −5.346 | −4.085 | −5.502 |
| 1st Quarter | −0.215 | −0.178 | −0.302 | −0.157 |
| Median | −0.013 | −0.008 | −0.024 | −0.010 |
| Mean | 0.000 | 0.000 | 0.000 | 0.000 |
| 3rd Quarter | 0.000 | 0.000 | 0.321 | 0.000 |
| Maximum | 6.481 | 5.674 | 5.009 | 7.644 |
| Skewness | 0.043 | 0.142 | −0.209 | 0.193 |
| Curtosis | 2.839 | 2.716 | 2.247 | 7.385 |
| *K–S* test statistic | 0.254 | 0.289 | 0.180 | 0.286 |
| *p*-value | $<0.001$ | $<0.001$ | $<0.001$ | $<0.001$ |

## Summary of findings, discussion, and conclusions

Solo research is a result of voluntarily made individual authorship decisions. Choosing solo research is as consequential for academic careers as choosing same-sex or mixed-sex collaborations, or choosing institutional, national, and international collaborations. Individual authorship decisions accumulate over time, accompanying academic careers. While it is known that authorship decisions need to be "intelligent" (Vafeas, 2010: 333) and "strategic" (Jeong et al., 2011: 968), men's decisions differ from women' decisions in different institutional types, disciplines, and national systems. Women have often been reported to be underrepresented as solo publishers (Sarsons et al., 2020; Walker, 2019; West et al., 2013). This research is the first to comprehensively study the gender solo research gap in the context of prior literature on solo research and solo research and gender within a large national higher education system: we examined the gap through "individual publication portfolios" constructed for each internationally visible Polish university professor ($N=25{,}463$, all assistant, associate, and full professors, and their 158,743 articles published in 2009–2018, including 18,900 solo articles). The population examined includes all university professors with at least a doctoral degree and with at least a single Scopus-indexed publication: from a slightly exclusionary, international perspective, we studied the whole research-active Polish academic profession.

Solo research is a special case of academic publishing where scientists, on the one hand, compete individually in the academic marketplace of ideas, taking full responsibility and full risk for publications' errors (Hudson, 1996; Kuld and O'Hagan, 2018), and on the other, where there is no ambiguity in credit allocation, and credits unambiguously go to the single author, sending clear signals about their research ability and independence (Barlow et al., 2017; Sarsons, 2017; Sarsons et al., 2020). Solo research has been expected to disappear for half a century (Price, 1963), but for many reasons, it continues to exist (West et al., 2013). In this paper, we examined "the gender solo research gap," existing alongside many other input-related gender gaps (e.g., the gender mobility or the gender international collaboration gaps) and output-related gender gaps in science (e.g., the gender tenure or the gender salary gap).

Our focus was on gender differences in solo research from a macro-level perspective of a single national higher education system: in our unique biographical, administrative, publication, and citation database ("Polish Science Observatory"), we have metadata on all scientists and on all their Scopus-indexed publications, including solo publications, in all academic disciplines. Our focus was on how male and female scientists of various disciplines and biological ages, employed in different institutional types make use of solo publishing.

Somehow surprisingly in the context of previous literature, while differences between male and female scientists in the individual publishing solo rate were in several cases statistically significant, they turned out to have no practical significance. We used the practical significance/statistical significance difference in our study, based on the effect-size *r* coefficient, following Leech et al. (2015: 196) comments that “a statistically significant result with a small effect size means that we can be very confident that there is *some* difference or association, but it is probably small and may not be practically important”. This is exactly the Polish case.

Across all disciplines, the individual publishing solo rate (or the share of solo articles in all articles published by a scientist and ranging from 0 to 1) was slightly higher for men than it was for women, and it was higher for men in most male-dominated disciplines. However, female scientists showed higher intensity of publishing solo (the rate for solo scientists only, or scientists with at least one solo paper in their individual publication portfolio) than male scientists did. The differences in the rate within disciplines by gender and between male-dominated and female-dominated disciplines by gender were much smaller than expected from previous literature. Cross-gender differences were actually more visible in the intensity of solo publishing by discipline than in individual publishing solo rates.

Previous literature indicated that the more research-focused an institution was, the higher involvement in publishing solo the scientist would have (e.g., Vafeas, 2010: 340). We contrasted the 10 research-intensive institutions involved in the Polish Excellence Initiative (IDUB) with 75 other institutions and found that the individual publishing solo rate for women in research-intensive institutions was higher than the rate for them elsewhere in the system. In research-intensive institutions, the intensity of solo publishing for women was substantially higher than it was for men by 10 percentage points, which was in accordance with previous literature. These gender differences were indeed statistically significant but their practical significance was marginal.

We hypothesized that younger scientists would publish solo articles significantly more often than older scientists would. However, surprisingly in the context of previous findings, the hypothesis found no support in our data. We examined the individual solo publishing rate by gender and age. Younger scientists had significantly lower individual solo publishing rates than middle-aged scientists did, and the differences between age groups were higher for women than for men. The rate for young male scientists was higher than that for young female scientists, again in line with previous studies.

However, the only gender difference in the solo rate with practical significance was between men and women under 40. The year-by-year approach confirmed similar trends for both genders. The intensity of solo publishing for female scientists (or the solo rate for scientists who ever published solo articles) was at least equal for each age; specifically, it was much higher for young female scientists under 40. Female solo scientists in their 30s emerged with a substantially higher share of solo articles in their publication portfolios: i.e. female scientists already involved in solo research were involved in solo research more intensively.

Finally, we used a multidimensional approach and our expectation was that being a female scientist would decrease the propensity to conduct solo research (we used a fractional logit regression model). Using a partial effects of fractional logistic regression approach, we estimated the probability of conducting solo research. The selected predictors of the individual publishing solo rate in the estimated models explained a high percentage of the variability of the dependent variable, from 77.5% in the model for associate professors to 82.3% in the model describing relationships in the population of full professors. However, most importantly, in none of the models did gender explain the variability of the rate. The strongest predictor was the average team size, that is, the number of co-authors. Publishing in STEM fields negatively affected the rate, publishing in male-dominated disciplines positively affected it, and the influence of international collaboration was negative. Finally, working in research-intensive universities resulted in an average rate increase of slightly more than 1 pp for all faculty except full professors.

In short, the gender solo research gap that emerges from our research is clearly much weaker than expected: within a more general trend in Polish science away from solo research and toward team research and away from national research and toward international research (Kwiek, 2020), gender differences in solo publishing seem to be less relevant than initially assumed based on the research literature. The larger context of the dominating team research in science overshadows the smaller context of gender differences in solo publishing. Our expectations of young female scientists conducting considerably less solo research than male scientists were not confirmed; there exists a gender solo research gap in Poland, but it is not wide and gender differences in solo publishing are of limited practical significance.

While research by Vafeas (2010) and Kuld and O'Hagan (2018) indicated a much higher role of solo research for young scientists in general, irrespective of their gender, our research does not confirm these findings in the Polish case: the highest share of solo research for both genders is noted for middle-aged scientists (40–54). Surprisingly, a bigger gender difference was noted in solo research intensity, not studied in previous literature, operationalized as the share of single-authored publications among all publications within individual publication portfolios of scientists with any solo research. Those women who are solo scientists use this mode of publishing more intensively than male solo scientists do (especially while finishing or just after their doctoral dissertations at the age of around 30–35).

Further research could include two new dimensions: a historical and a global one. A new research question could be how the changing shares of solo research in individual publication portfolios by gender evolve over time in Poland and evolve globally. Specifically, we could ask whether the changes in the individual solo publishing rates over time and from a cross-country perspective are similar for both male and female scientists and whether the gender solo research gap is widening or closing in individual disciplines. Our current "Polish Science Observatory" database includes publications from the decade of 2009–2018, and the rate from this period could be compared with rates in the previous decades (e.g., the 2000s and 1990s) to examine the temporal dynamics of changing publishing patterns. The same temporal limitation to a single decade pertains to our parallel "OECD Science Observatory" database of all (gender-defined) scientists and all (gender-classified) articles indexed in Scopus from 1674 research-involved institutions in 38 OECD economies from the same period; in addition, in this possible cross-national comparative study, biological age would need to be replaced with the academic age, or the time that has passed since the first publication (as in Kuld & O'Hagan, 2018; Robinson-Garcia et al., 2020). On top of this, certainly, academic careers and the gender gaps that accompany them can be more meaningfully studied if publication and citation data are combined not only with administrative and biographical data, as in this research, but also with large-scale surveys of the academic profession which we are planning to conduct.

## Appendix

See Tables 10 and 11.

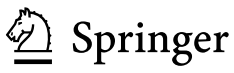

**Table 10** Structure of the sample, all Polish internationally visible university professors, by gender, age group, academic position, and discipline (by discipline type: STEM and non-STEM, female-dominated and male-dominated), presented with column and row percentages (young scientists: aged 39 years and younger; middle-aged: aged 40–54 years; and older, aged 55 and more)

| | Female | | | Male | | | Total | | |
|---|---|---|---|---|---|---|---|---|---|
| | *N* | % Col | % Row | *N* | % Col | % Row | *N* | % Col | % Row |
| *Age group* | | | | | | | | | |
| Young | 3128 | 29.6 | 49.4 | 3199 | 21.5 | 50.6 | 6327 | 24.8 | 100.0 |
| Middle-aged | 5584 | 52.8 | 44.1 | 7074 | 47.5 | 55.9 | 12,658 | 49.7 | 100.0 |
| Older | 1865 | 17.6 | 28.8 | 4613 | 31.0 | 71.2 | 6478 | 25.4 | 100.0 |
| Total | 10,577 | 100.0 | 41.5 | 14,886 | 100.0 | 58.5 | 25,463 | 100.0 | 100.0 |
| *Academic position* | | | | | | | | | |
| Assistant prof | 6851 | 64.8 | 48.0 | 7420 | 49.8 | 52.0 | 14,271 | 56.0 | 100.0 |
| Asssoc. prof | 2822 | 26.7 | 38.0 | 4596 | 30.9 | 62.0 | 7418 | 29.1 | 100.0 |
| Full professor | 904 | 8.5 | 24.0 | 2870 | 19.3 | 76.0 | 3774 | 14.8 | 100.0 |
| Total | 10,577 | 100.0 | 41.5 | 14,886 | 100.0 | 58.5 | 25,463 | 100.0 | 100.0 |
| *Discipline (ASJC)—STEM* | | | | | | | | | |
| AGRI | 1444 | 13.7 | 53.4 | 1258 | 8.5 | 46.6 | 2702 | 10.6 | 100.0 |
| BIO | 1068 | 10.1 | 60.0 | 712 | 4.8 | 40.0 | 1780 | 7.0 | 100.0 |
| CHEM | 756 | 7.1 | 51.3 | 719 | 4.8 | 48.7 | 1475 | 5.8 | 100.0 |
| CHEMENG | 185 | 1.7 | 38.5 | 296 | 2.0 | 61.5 | 481 | 1.9 | 100.0 |
| COMP | 170 | 1.6 | 16.5 | 860 | 5.8 | 83.5 | 1030 | 4.0 | 100.0 |
| DEC | 24 | 0.2 | 44.4 | 30 | 0.2 | 55.6 | 54 | 0.2 | 100.0 |
| EARTH | 385 | 3.6 | 33.4 | 769 | 5.2 | 66.6 | 1154 | 4.5 | 100.0 |
| ENER | 82 | 0.8 | 27.8 | 213 | 1.4 | 72.2 | 295 | 1.2 | 100.0 |
| ENG | 501 | 4.7 | 14.9 | 2857 | 19.2 | 85.1 | 3358 | 13.2 | 100.0 |
| ENVIR | 848 | 8.0 | 50.5 | 832 | 5.6 | 49.5 | 1680 | 6.6 | 100.0 |
| IMMU | 90 | 0.9 | 75.6 | 29 | 0.2 | 24.4 | 119 | 0.5 | 100.0 |
| MATER | 495 | 4.7 | 33.9 | 967 | 6.5 | 66.1 | 1462 | 5.7 | 100.0 |
| MATH | 259 | 2.4 | 25.2 | 767 | 5.2 | 74.8 | 1026 | 4.0 | 100.0 |
| PHARM | 169 | 1.6 | 66.5 | 85 | 0.6 | 33.5 | 254 | 1.0 | 100.0 |
| PHYS | 182 | 1.7 | 16.6 | 916 | 6.2 | 83.4 | 1098 | 4.3 | 100.0 |
| *Discipline (ASJC)—non-STEM* | | | | | | | | | |
| BUS | 372 | 3.5 | 52.1 | 342 | 2.3 | 47.9 | 714 | 2.8 | 100.0 |
| DENT | 57 | 0.5 | 76.0 | 18 | 0.1 | 24.0 | 75 | 0.3 | 100.0 |
| ECON | 186 | 1.8 | 49.1 | 193 | 1.3 | 50.9 | 379 | 1.5 | 100.0 |
| HEALTH | 23 | 0.2 | 34.3 | 44 | 0.3 | 65.7 | 67 | 0.3 | 100.0 |
| HUM | 527 | 5.0 | 49.8 | 531 | 3.6 | 50.2 | 1058 | 4.2 | 100.0 |
| MED | 1920 | 18.2 | 53.7 | 1654 | 11.1 | 46.3 | 3574 | 14.0 | 100.0 |
| PSYCH | 194 | 1.8 | 63.8 | 110 | 0.7 | 36.2 | 304 | 1.2 | 100.0 |
| SOC | 494 | 4.7 | 49.8 | 498 | 3.3 | 50.2 | 992 | 3.9 | 100.0 |
| VET | 146 | 1.4 | 44.0 | 186 | 1.2 | 56.0 | 332 | 1.3 | 100.0 |
| Total | 10,577 | 100.0 | 41.5 | 14,886 | 100.0 | 58.5 | 25,463 | 100.0 | 100.0 |
| *Gender domination in discipline* | | | | | | | | | |
| Female-dom | 6918 | 65.4 | 54.6 | 5759 | 38.7 | 45.4 | 12,677 | 49.8 | 100.0 |
| Male-dom | 3659 | 34.6 | 28.6 | 9127 | 61.3 | 71.4 | 12,786 | 50.2 | 100.0 |
| Total | 10,577 | 100.0 | 41.5 | 14,886 | 100.0 | 58.5 | 25,463 | 100.0 | 100.0 |
| *Mean publication prestige (percentile)* | | | | | | | | | |
| <0,30) | 777 | 7.3 | 48.7 | 817 | 5.5 | 51.3 | 1594 | 6.3 | 100.0 |

**Table 10** (continued)

| | Female | | | Male | | | Total | | |
|---|---|---|---|---|---|---|---|---|---|
| | *N* | % Col | % Row | *N* | % Col | % Row | *N* | % Col | % Row |
| <30,40) | 888 | 8.4 | 41.3 | 1262 | 8.5 | 58.7 | 2150 | 8.4 | 100.0 |
| <40,50) | 1432 | 13.5 | 39.7 | 2171 | 14.6 | 60.3 | 3603 | 14.1 | 100.0 |
| <50,60) | 2778 | 26.3 | 40.8 | 4023 | 27.0 | 59.2 | 6801 | 26.7 | 100.0 |
| <60,70) | 2573 | 24.3 | 40.8 | 3728 | 25.0 | 59.2 | 6301 | 24.7 | 100.0 |
| <70,80) | 1691 | 16.0 | 43.4 | 2202 | 14.8 | 56.6 | 3893 | 15.3 | 100.0 |
| <80,90) | 373 | 3.5 | 39.4 | 573 | 3.8 | 60.6 | 946 | 3.7 | 100.0 |
| <90,100) | 65 | 0.6 | 37.1 | 110 | 0.7 | 62.9 | 175 | 0.7 | 100.0 |
| Total | 10,577 | 100.0 | 41.5 | 14,886 | 100.0 | 58.5 | 25,463 | 100.0 | 100.0 |

**Table 11** Changes of the mean individual publishing solo rate (range: 0–1) by discipline over time (2009–2018), all scientists in the sample ($N = 25,463$ scientists)

| Discipline | 2009 | 2010 | 2011 | 2012 | 2013 | 2014 | 2015 | 2016 | 2017 | 2018 |
|---|---|---|---|---|---|---|---|---|---|---|
| AGRI | 0.0848 | 0.0822 | 0.0655 | 0.0532 | 0.0514 | 0.0380 | 0.0201 | 0.0305 | 0.0235 | 0.0218 |
| BIO | 0.0140 | 0.0154 | 0.0162 | 0.0124 | 0.0127 | 0.0105 | 0.0094 | 0.0075 | 0.0078 | 0.0063 |
| BUS | 0.4615 | 0.3490 | 0.3070 | 0.3509 | 0.3171 | 0.3018 | 0.2435 | 0.2279 | 0.2174 | 0.2154 |
| CHEM | 0.0409 | 0.0432 | 0.0379 | 0.0327 | 0.0339 | 0.0243 | 0.0253 | 0.0278 | 0.0248 | 0.0224 |
| CHEMENG | 0.0790 | 0.1237 | 0.0977 | 0.0973 | 0.0588 | 0.0746 | 0.0629 | 0.0401 | 0.0677 | 0.0644 |
| COMP | 0.1670 | 0.1577 | 0.1518 | 0.1684 | 0.1319 | 0.1517 | 0.1145 | 0.1307 | 0.1141 | 0.1201 |
| DEC | 0.4286 | 0.5000 | 0.2000 | 0.5000 | 0.5741 | 0.6818 | 0.3125 | 0.3261 | 0.3913 | 0.2059 |
| EARTH | 0.1770 | 0.1930 | 0.1790 | 0.1507 | 0.1436 | 0.1135 | 0.1226 | 0.1309 | 0.0994 | 0.0826 |
| ECON | 0.4483 | 0.4434 | 0.4353 | 0.3284 | 0.3903 | 0.3795 | 0.2789 | 0.2770 | 0.2510 | 0.2719 |
| ENER | 0.1481 | 0.1828 | 0.2368 | 0.1873 | 0.1167 | 0.1156 | 0.1411 | 0.1270 | 0.1343 | 0.1287 |
| ENG | 0.1951 | 0.2041 | 0.1788 | 0.1812 | 0.1539 | 0.1612 | 0.1225 | 0.1206 | 0.1100 | 0.0993 |
| ENVIR | 0.0981 | 0.0901 | 0.0886 | 0.1019 | 0.0846 | 0.0674 | 0.0445 | 0.0517 | 0.0470 | 0.0462 |
| HEALTH | 0.1333 | 0.1250 | 0.0833 | 0.0333 | 0.0690 | 0.0549 | 0.0926 | 0.1609 | 0.0373 | 0.0575 |
| HUM | 0.7880 | 0.8108 | 0.8018 | 0.7210 | 0.7399 | 0.6773 | 0.7577 | 0.7082 | 0.7660 | 0.6131 |
| IMMU | 0.0094 | 0.0051 | 0.0258 | 0.0152 | 0.0163 | 0.0238 | 0.0127 | 0.0086 | 0.0171 | 0.0000 |
| MATER | 0.0620 | 0.0468 | 0.0578 | 0.0604 | 0.0657 | 0.0517 | 0.0367 | 0.0338 | 0.0399 | 0.0409 |
| MATH | 0.3201 | 0.2646 | 0.2760 | 0.2628 | 0.2711 | 0.2462 | 0.2480 | 0.2245 | 0.2071 | 0.1685 |
| MED | 0.0155 | 0.0158 | 0.0151 | 0.0168 | 0.0138 | 0.0144 | 0.0137 | 0.0125 | 0.0089 | 0.0090 |
| NURS | 0.0000 | 0.0000 | 0.0000 | n.a | 0.0000 | 0.1667 | 0.0000 | 0.4000 | 0.0000 | 0.0000 |
| PHARM | 0.0228 | 0.0131 | 0.0240 | 0.0147 | 0.0069 | 0.0064 | 0.0036 | 0.0019 | 0.0028 | 0.0070 |
| PHYS | 0.0768 | 0.0680 | 0.0914 | 0.0606 | 0.0575 | 0.0551 | 0.0549 | 0.0461 | 0.0497 | 0.0453 |
| PSYCH | 0.2551 | 0.2969 | 0.1534 | 0.1563 | 0.0936 | 0.2108 | 0.1056 | 0.1511 | 0.1069 | 0.1177 |
| SOC | 0.6740 | 0.5994 | 0.6656 | 0.4827 | 0.5611 | 0.4729 | 0.3934 | 0.4278 | 0.3179 | 0.3725 |
| VET | 0.0202 | 0.0266 | 0.0135 | 0.0265 | 0.0147 | 0.0133 | 0.0140 | 0.0045 | 0.0031 | 0.0033 |

**Acknowledgements** We gratefully acknowledge the support of the Ministry of Education and Science through its Dialogue Grant 0022/DLG/2019/10 (RESEARCH UNIVERSITIES).

## References

Abramo, G., Aksnes, D. W., & D'Angelo, C. A. (2020). Comparison of research productivity of Italian and Norwegian professors and universities. *Journal of Informetrics, 14*(2), 101023.

Abramo, G., D'Angelo, C. A., & Di Costa, F. (2019). A gender analysis of top scientists' collaboration behavior: Evidence from Italy. *Scientometrics, 120*, 405–418.

Abramo, G., D'Angelo, C. A., & Murgia, G. (2013). Gender differences in research collaboration. *Journal of Informetrics, 7*, 811–822.

Abramo, G., D'Angelo, C. A., & Rosati, F. (2015). Selection committees for academic recruitment: Does gender matter? *Research Evauation, 24*(4), 392–404.

Abt, H. A. (2007). The future of single-authored papers. *Scientometrics, 73*, 353–358. https://doi.org/10.1007/s11192-007-1822-9

Ackers, L. (2008). Internationalization, mobility, and metrics: A new form of indirect discrimination? *Minerva, 46*, 411–435.

Aksnes, D. W., Piro, F. N., & Rørstad, K. (2019). Gender gaps in international research collaboration: A bibliometric approach. *Scientometrics, 120*, 747–774.

Aksnes, D. W., Rørstad, K., Piro, F. N., & Sivertsen, G. (2011). Are female researchers less cited? A large scale study of Norwegian researchers. *J. Am. Soc. Inf. Sci. Tech., 62*(4), 628–636.

Allen, L., Scott, J., Brand, A., Hlava, M., & Altman, M. (2014). Credit where credit is due. *Nature, 508*(7496), 312–313.

Avolio, B., Chávez, J., & Vílchez-Román, C. (2020). Factors that contribute to the underrepresentation of women in science careers worldwide: A literature review. *Social Psychology of Education., 23*, 773–794.

Barbezat, D. A., & Hughes, J. W. (2005). Salary structure effects and the gender pay gap in academia. *Research in Higher Education, 46*(6), 621–640.

Barlow, J., Stephens, P. A., Bode, M., Cadotte, M. W., Lucas, K., Newton, E., Nuñez, M. A., & Pettorelli, N. (2017). On the extinction of the single-authored paper: The causes and consequences of increasingly collaborative applied ecological research. *Journal of Applied Ecology, 55*(1), 1–4.

Blau, F. D., & Kahn, L. M. (2000). Gender differences in pay. *Journal of Economic Perspectives, 14*, 75–99.

Bozeman, B., Fay, D., & Slade, C. P. (2012). Research collaboration in universities and academic entrepreneurship: The-state-of-the-art. *The Journal of Technology Transfer, 38*(1), 1–67.

Bridgstock, M. (1991). The quality of single and multiple authored papers: An unresolved problem. *Scientometrics, 21*(1), 37–48.

Ceci, S. J., Ginther, D. K., Kahn, S., & Williams, W. M. (2014). Women in academic science: A changing landscape. *Psychological Science in the Public Interest, 15*(3), 75–141.

Ceci, S. J., & Williams, W. M. (2011). Understanding current causes of women's underrepresentation in science. *Proceedings of the National Academy of Sciences, 108*(8), 3157–3162.

Clauset, A., Arbesman, S., & Larremore, D. B. (2015). Systematic inequality and hierarchy in faculty hiring networks. *Science Advances, 1*(1), e1400005–e1400005.

Cohen, J. (1988). *Statistical power and analysis for the behavioral sciences*. Lawrence Erlbaum.

Cole, J. R. (1979). *Fair science: Women in the scientific community*. Columbia University Press.

Cruz-Castro, L., & Sanz-Menéndez, L. (2019) Grant allocation disparities from a gender perspective: Literature review. Synthesis Report. GRANteD Project D.1.1. https://doi.org/10.20350/digitalCSIC/10548.

Cummings, J. N., & Kiesler, S. (2007). Coordination costs and project outcomes in multi-university collaborations. *Research Policy., 36*, 1620–1634.

Cummings, W. K., & Finkelstein, M. J. (2012). Conclusion: New rules and roles. In W. K. Cummings & M. J. Finkelstein (Eds.), *Scholars in the changing American academy* (pp. 141–151). Springer. https://doi.org/10.1007/978-94-007-2730-4_10

De Solla Price, D. J. (1963). *Little science, big science*. Columbia University Press. https://doi.org/10.7312/pric91844

Diezmann, C., & Grieshaber, S. (2019). *Women professors. Who makes it and how?* Springer Nature.

Enamorado, T., Fifield, B., & Imai, K. (2019). Using a probabilistic model to assist merging of large-scale administrative records. *American Political Science Review, 113*(2), 353–371.

Endersby, J. W. (1996). Collaborative research in the social sciences: Multiple authorship and publication credit. *Social Science Quarterly, 77*, 375–392.

Farber, M. (2005). Single-authored publications in the sciences at Israeli universities. Journal of Information Science, 31(1), 62–66.

Feeney, M. K., & Bernal, M. (2010). Women in STEM networks: Who seeks advice and support from women scientists? *Scientometrics, 85*(3), 767–790.

Fell, C. B., & König, C. J. (2016). Is there a gender difference in scientific collaboration? A scientometric examination of co-authorships among industrial-organizational psychologists. *Scientometrics, 108*(1), 113–141.

Field, A., Miles, J. and Field, Z. (2012). Discovering statistics using R. Sage Publications Ltd., London.

Fisher, B. S., Cobane, C. T., Ven, T. M. V., & Cullen, F. T. (1998). How many authors does it take to publish an article? Trends and patterns in political science. *PS: Political Science and Politics, 31*(4), 847–856.

Fochler, M., Felt, U., & Müller, R. (2016). Unsustainable growth, hyper-competition, and worth in life science research: Narrowing evaluative repertoires in doctoral and postdoctoral scientists' work and lives. *Minerva, 54*(2), 175–200.

Fox, M. F. (1985). Location, sex-typing, and salary among academics. *Work and Occupations, 12*(2), 186–205.

Fox, M. F., Realff, M. L., Rueda, D. R., & Morn, J. (2017). International research collaboration among women engineers: Frequency and perceived barriers, by regions. *Journal of Technology Transfer, 42*(6), 1292–1306.

Fox, M. F. (2020). Gender science and academic rank: Key issues and approaches. *Quantitative Science Studies, 1*(3), 1001–1006. https://doi.org/10.1162/qss_a_00057

Frehill, L. M., & Zippel, K. (2006). (2010) Gender and international collaborations of academic scientists and engineers: Findings from the survey of doctorate recipients. *Journal of the Washington Academy of Sciences, 97*(1), 49–69.

Georghiou, L. (1998). Global cooperation in research. *Research Policy, 27*, 611–628.

Ghiasi, G, Mongeon, P., Sugimoto, C., & Larivière, V. (2018) Gender homophily in citations. In *Conference Proceedings: the 3rd International Conference on Science and Technology Indicators (STI 2018)* (pp. 1519–1525).

Ghiasi, G., Sainte-Marie, M., & Larivière, V. (2019) Making it personal: Examining personalization patterns of single-authored papers. In *17th International Conference on Scientometrics & Informetrics. September 2–5, 2019* (pp. 2088–2093).

Ghiasi, G., Larivière, V., & Sugimoto, C. R. (2015). On the compliance of women engineers with a gendered scientific system. *PLoS ONE, 10*(12), 1–19.

Glänzel, W. (2002). Coauthorship patterns and trends in the sciences 1980–1998: A bibliometric study with implications for database indexing and search strategies. *Library Trends, 50*(3), 461–473.

Goastellec, G., & Vaira, M. (2017). Women's place in academia: Case studies of Italy and Switzerland. In H. Eggins (Ed.), *The changing role of women in higher education* (pp. 173–191). Springer.

Greguletz, E., Diehl, M.-R., & Kreutzer, K. (2018). Why women build less effective networks than men: The role of structural exclusion and personal hesitation. *Human Relations, 72*, 1234.

Gupta, N. D., Poulsen, A., & Villeval, M. C. (2013). Gender matching and competitiveness: Experimental evidence. *Economic Inquiry, 51*(1), 816–835.

Halevi, G. (2019). Bibliometric studies on gender disparities in science. In W. Glänzel, H. F. Moed, U. Schmoch, & M. Thelwall (Eds.), *Springer handbook of science and technology indicators* (pp. 563–580). Springer.

Hartley, J. (2005). Refereeing and the single author. *Journal of Information Science, 31*(3), 251–256. https://doi.org/10.1177/0165551505052474

Heffernan, T. (2020). Academic networks and career trajectory: There's no career in academia without networks. *Higher Education Research & Development*. https://doi.org/10.1080/07294360.2020.1799948

Heffner, A. G. (1979). Authorship recognition of subordinates in collaborative research. *Social Studies of Science, 9*(3), 377–384.

Henriksen, D. (2016). The rise in co-authorship in the social sciences (1980–2013). *Scientometrics, 107*, 455–476.
Herzog, T. N., Scheuren, F. J., & Winkler, W. E. (2007). *Data quality and record linkage techniques*. Springer.
Holman, L., & Morandin, C. (2019). Researchers collaborate with same-gendered colleagues more often than expected across the life sciences. *PLOS ONE, 14*(4), e0216128.
Huang, D.-W. (2015). Temporal evolution of multi-author papers in basic sciences from 1960 to 2010. *Scientometrics, 105*, 2137–2147.
Huang, J., Gates, A. J., Sinatra, R., & Barabási, A.-L. (2020). Historical comparison of gender inequality in scientific careers across countries and disciplines. *Proceedings of the National Academy of Sciences of the United States of America, 117*(9), 4609–4616.
Hudson, J. (1996). Trends in multi-authored papers in economics. *The Journal of Economic Perspectives: A Journal of the American Economic Association, 10*(3), 153–158.
Hutson, S. R. (2006). Self-citation in archaeology: Age, gender, prestige, and the self. *Journal of Archaeological Method and Theory, 13*(1), 1–18.
Jabbehdari, S., & Walsh, J. P. (2017). Authorship norms and project structures in science. *Science, Technology, & Human Values, 42*(5), 872–900.
Jadidi, M., Karimi, F., Lietz, H., & Wagner, C. (2018). Gender disparities in science? Dropout, productivity, collaborations, and success of male and female computer scientists. *Advances in Complex Systems, 21*(3–4), 1750011.
Jeong, S., Choi, J. Y., & Kim, J. (2011). The determinants of research collaboration modes: Exploring the effects of research and researcher characteristics on co-authorship. *Scientometrics, 89*(3), 967–983.
Jeong, S., Choi, J. Y., & Kim, J. Y. (2014). On the drivers of international collaboration: The impact of informal communication, motivation, and research resources. *Science and Public Policy, 41*(4), 520–531.
Jöns, H. (2011). Transnational academic mobility and gender. *Globalisation, Societies and Education, 9*(2), 183–209.
Kanter, R. M. (1977). Some effects of proportions on group life: Skewed sex ratios and responses to token women. *American Journal of Sociology, 82*(5), 965–990.
Kegen, N. V. (2013). Science networks in cutting-edge research institutions: Gender homophily and embeddedness in formal and informal networks. *Procedia: Social and Behavioral Sciences, 79*, 62–81.
Key, E., & Sumner, J. L. (2019). You research like a girl: Gendered research agendas and their implications. *PS: Political Science & Politics, 52*(4), 663–668.
King, M. M., Bergstrom, C. T., Correll, S. J., Jacquet, J., & West, J. D. (2017). Men set their own cites high: Gender and self-citation across fields and over time. *Socius, 3*, 1–22.
Kuld, L., & O'Hagan, J. (2018). Rise of multi-authored papers in economics: Demise of the 'lone star' and why? *Scientometrics, 114*, 1207–1225.
Kwiek, M. (2015). The internationalization of research in Europe: A quantitative study of 11 national systems from a micro-level perspective. *Journal of Studies in International Education, 19*(2), 341–359.
Kwiek, M. (2016). The European research elite: A cross-national study of highly productive academics across 11 European systems. *High Educ., 71*(3), 379–397.
Kwiek, M. (2018a). Academic top earners: Research productivity, prestige generation and salary patterns in European universities. *Science and Public Policy., 45*(1), 1–13.
Kwiek, M. (2018b). High research productivity in vertically undifferentiated higher education systems: Who are the top performers? *Scientometrics, 115*(1), 415–462.
Kwiek, M. (2019). *Changing European academics. A comparative study of social stratification, work patterns and research productivity*. London and New York: Routledge.
Kwiek, M. (2020). Internationalists and locals: International research collaboration in a resource-poor system. *Scientometrics, 124*, 57–105.
Kwiek, M. (2021a). The prestige economy of higher education journals: A quantitative approach. *Higher Education, 81*, 493–519.
Kwiek, M. (2021b). What large-scale publication and citation data tell us about international research collaboration in Europe: Changing national patterns in global contexts. *Studies in Higher Education, 46*(12), 2629–2649.
Kwiek, M., & Roszka, W. (2021a). Gender disparities in international research collaboration: A large-scale bibliometric study of 25,000 university professors. *Journal of Economic Surveys, 35*(5), 1344–1380.
Kwiek, M., & Roszka, W. (2021b). Gender-based homophily in research: A large-scale study of man-woman collaboration. *Journal of Informetrics, 15*(3), 1–38.
Landry, R., & Amara, N. (1998). The impact of transaction costs on the institutional structuration of collaborative academic research. *Research Policy., 27*, 901–913.

Larivière, V., & Gingras, Y. (2010). The impact factor's Matthew effect: A natural experiment in bibliometrics. *Journal of the American Society for Information Science and Technology., 61*(2), 424–427.

Larivière, V., Gingras, Y., & Archambault, É. (2006). Canadian collaboration networks: A comparative analysis of the natural sciences, social sciences and the humanities. *Scientometrics, 68*(3), 519–533. https://doi.org/10.1007/s11192-006-0127-8

Larivière, V., Sugimoto, C. R., Chaoquin, N., Gingras, Y., & Cronin, B. (2013). Global gender disparities in science. *Nature, 504*, 211–213.

Larivière, V., Sugimoto, C. R., Tsou, A., & Gingras, Y. (2015). Team size matters: Collaboration and scientific impact since 1900. *Journal of the Association for Information Science and Technology, 66*(7), 1323–1332.

Larivière, V., Vignola-Gagné, E., Villeneuve, C., Gelinas, P., & Gingras, Y. (2011). Sex differences in research funding, productivity and impact: An analysis of Quebec university professors. *Scientometrics, 87*(3), 483–498.

Leech, N. L., Barret, K. C., & Morgan, G. A. (2015). *IBM SPSS for Intermediate Statistics. Use and interpretation*. Routledge Taylor & Francis Group.

Leišytė, L., & Hosch-Dayican, B. (2017). Gender and academic Work at a Dutch university. In H. Eggins (Ed.), *The changing role of women in higher education* (pp. 95–117). Springer.

Lerchenmueller, M., Hoisl, K., & Schmallenbach, L. (2019). Homophily, biased attention, and the gender gap in science. *Academy of Management Proceedings, 2019*(1), 14784. https://doi.org/10.5465/AMBPP.2019.14784abstract

Lindsay, L. (2011). *Gender roles: A sociological perspective* (5th ed.). Prentice Hall.

Maddi, A., Larivière, V., & Gingras, Y. (2019) Man-woman collaboration behaviors and scientific visibility: Does gender affect the academic impact in economics and management? In *Proceedings of the 17th International Conference on Scientometrics & Informetrics, September 2–5, 2019* (pp. 1687–1697).

Madison, G., & Fahlman, P. (2020). Sex differences in the number of scientific publications and citations when attaining the rank of professor in Sweden. *Studies in Higher Education*. https://doi.org/10.1080/03075079.2020.1723533

Maliniak, D., Powers, R., & Walter, B. F. (2013). The gender citation gap in international relations. *International Organization, 67*(4), 889–922.

Marsh, H. W., Bornmann, L., Mutz, R., Daniel, H.-D., & O'Mara, A. (2009). Gender effects in the peer reviews of grant proposals: A comprehensive meta-analysis comparing traditional and multilevel approaches. *Review of Educational Research, 79*(3), 1290–1326.

McDowell, J. M., & Smith, J. K. (1992). The effect of gender-sorting on propensity to coauthor: Implications for academic promotion. *Economic Inquiry, 30*(1), 68–82.

McDowell, M. J., Singell, L. D., Jr., & Stater, M. (2006). Two to tango? Gender differences in the decisions to publish and coauthor. *Economic Inquiry, 44*(1), 153–168.

Merton, R. K. (1968). The Matthew effect in science. *Science, 159*(3810), 56–63.

Mihaljević-Brandt, H., Santamaría, L., & Tullney, M. (2016). The effect of gender in the publication patterns in mathematics. *PLOS ONE, 11*(10), e0165367.

Miller, J., & Chamberlin, M. (2000). Women are teachers, men are professors: A study of student perceptions. *Teaching Sociology, 28*(4), 283–298.

Mishra, S., Fegley, B. D., Diesner, J., & Torvik, V. I. (2018). Self-citation is the hallmark of productive authors, of any gender. *PLOS ONE, 13*(9), e0195773.

Müller, R. (2012). Collaborating in life science research groups: The question of authorship. *Higher Education Policy, 25*(3), 289–311. https://doi.org/10.1057/hep.2012.11

Müller, R., & Kenney, M. (2014). Agential conversations: Interviewing postdoctoral life scientists and the politics of mundane research practices. *Science as Culture, 23*(4), 537–559. https://doi.org/10.1080/09505431.2014.916670

Nabout, J. C., Parreira, M. R., Teresa, F. B., Carneiro, F. M., da Cunha, H. F., de Souca Ondei, L., et al. (2015). Publish (in a group) or perish (alone): The trend from single- to multi-authorship in biological papers. *Scientometrics, 102*, 357–364.

Nielsen, M. W. (2016). Gender inequality and research performance: Moving beyond individual-meritocratic explanations of academic advancement. *Studies in Higher Education, 41*(11), 2044–2060.

Olechnicka, A., Ploszaj, A., & Celinska-Janowicz, D. (2019). *The geography of scientific collaboration*. Routledge.

Papke, L. E., & Wooldridge, J. M. (1996). Econometric methods for fractional response variables with an application to 401(k) plan participation rates. *Journal of Applied Econometrics, 11*, 619–632.

Potthoff, M., & Zimmermann, F. (2017). Is there a gender-based fragmentation of communication science? An investigation of the reasons for the apparent gender homophily in citations. *Scientometrics, 112*(2), 1047–1063.

Ramalho, E. A., Ramalho, J. J. S., & Murteira, J. M. R. (2011). Alternative estimating and testing empirical strategies for fractional regression models. *Journal of Economic Surveys., 25*(1), 19–68.
Rivera, L. A. (2017). When two bodies are (not) a problem: Gender and relationship status discrimination in academic hiring. *American Sociological Review., 82*, 1111–1138.
Robinson-Garcia, N., Costas, R., Sugimoto, C. R., Larivière, V., & Nane, G. F. (2020). Task specialization across research careers. *Elife, 2020*(9), e60586. https://doi.org/10.7554/eLife.60586
Rutledge, R., & Karim, K. (2009). Determinants of coauthorship for the most productive authors of accounting literature. *Journal of Education for Business, 84*(3), 130–134. https://doi.org/10.3200/joeb.84.3.130-134
Ryu, B. K. (2020) The demise of single-authored publications in computer science: A citation network analysis. arXiv: https://arxiv.org/abs/2001.00350.
Santos, J. M., Horta, H., & Amâncio, L. (2020). Research agendas of female and male academics: A new perspective on gender disparities in academia. *Gender and Education, 33*, 625.
Sarsons, H. (2017). Recognition for group work: Gender differences in academia. *American Economic Review, 107*(5), 141–145. https://doi.org/10.1257/aer.p20171126
Sarsons, H., Gërxhani, K., Reuben, E., & Schram, A. (2021). Gender differences in recognition for group work. *Journal of Political Economy, 129*(1), 101–147. https://doi.org/10.1086/711401
Scopus (2021) The Scopus dataset, available from www.scopus.com (institutional subscription required).
Shapin, S. (1991). "The mind is its own place": Science and solitude in seventeenth-century England. *Science in Context, 4*(1), 191–218.
Shapiro, J. R., & Williams, A. M. (2011). The role of stereotype threats in undermining girls' and women's performance and interest in STEM fields. *Sex Roles, 66*(3–4), 175–183.
Sitzmann, T., & Campbell, E. M. (2021). The hidden cost of prayer: Religiosity and the gender wage gap. *Academy of Management Journal*. https://doi.org/10.5465/amj.2019.1254
Sonnert, G., & Holton, G. (1996). Career patterns of women and men in the sciences. *American Scientist, 84*(1), 63–71.
Statistics Poland. (2021). *Higher education institutions and their finances in 2020*. Statistics Poland.
Stephan, P. (2012). *How economics shapes science*. Harvard University Press.
Sugimoto, C. R., Ni, C., & Larivière, V. (2015). On the relationship between gender disparities in scholarly communication and country-level development indicators. *Science and Public Policy*. https://doi.org/10.1093/scipol/scv007
Thelwall, M. (2020). Gender differences in citation impact for 27 fields and six English-speaking countries 1996–2014. *Quantitative Science Studies, 1*(2), 599–617.
Thelwall, M., Bailey, C., Tobin, C., & Bradshaw, N.-A. (2019). Gender differences in research areas, methods and topics: Can people and thing orientations explain the results? *Journal of Informetrics, 13*(1), 149–169.
Toutkoushian, R. K., & Bellas, M. L. (1999). Faculty time allocations and research productivity: Gender, race and family effects. *The Review of Higher Education, 22*(4), 367–390.
Uhly, K. M., Visser, L. M., & Zippel, K. S. (2017). Gendered patterns in international research collaborations in academia. *Studies in Higher Education, 42*(4), 760–782.
Vabø, A., Padilla-Gonzales, L. E., Waagene, E., & Naess, T. (2014). Gender and faculty internationalization. In F. Huang, M. Finkelstein, & M. Rostan (Eds.), *The internationalization of the academ changes, realities and prospects* (pp. 183–206). Springer.
Vafeas, N. (2010). Determinants of single authorship. *EuroMed Journal of Business, 5*(3), 332–344.
Van den Besselaar, P., & Sandström, U. (2015). Early career grants, performance, and careers: A study on predictive validity of grant decisions. *J. Informetr., 9*(4), 826–838.
Van den Besselaar, P., & Sandström, U. (2016). Gender differences in research performance and its impact on careers: A longitudinal case study. *Scientometrics, 106*(1), 143–162.
Van den Besselaar, P., & Sandström, U. (2017). Vicious circles of gender bias, lower positions, and lower performance: Gender differences in scholarly productivity and impact. *PLoS ONE*. https://doi.org/10.1371/journal.pone.0183301
Van den Brink, M., & Benschop, Y. (2013). Gender in academic networking: The role of gatekeepers in professorial recruitment. *Journal of Management Studies, 51*(3), 460–492.
Walker, K. A. (2019). Females are first authors, sole authors, and reviewers of entomology publications significantly less often than males. *Annals of the Entomological Society of America*. https://doi.org/10.1093/aesa/saz066
Wang, D., & Barabási, A. (2021). *The science of science*. Cambridge: Cambridge University Press.
Ward, M. E., & Sloane, P. J. (2000). Non-pecuniary advantages versus pecuniary disadvantages: Job satisfaction among male and female academics in Scottish universities. *Scottish Journal of Political Economy, 47*(3), 273–303.

Weisshaar, K. (2017). Publish and perish? An assessment of gender gaps in promotion to tenure in academia. *Social Forces, 96*(2), 529–560. https://doi.org/10.1093/sf/sox052
West, J. D., Jacquet, J., King, M. M., Correll, S. J., & Bergstrom, C. T. (2013). The role of gender in scholarly authorship. *PLOS ONE, 8*(7), e66212.
Wuchty, S., Jones, B. F., & Uzzi, B. (2007). The increasing dominance of teams in production of knowledge. *Science, 316*(5827), 1036–1039. https://doi.org/10.1126/science.1136099
Xie, Y., & Shauman, K. A. (2003). *Women in science: Career processes and outcomes*. Harvard University Press.
Zippel, K. (2017). *Women in global science*. Stanford University Press.